\documentclass[aps,11pt,prd,notitlepage,tightenlines,nofootinbib,superscriptaddress]{revtex4-2}
\usepackage{amsmath, amssymb, amsthm}
\usepackage{mathtools}
\usepackage{mathrsfs}
\usepackage{bm}
\usepackage{slashed}   
\usepackage{graphicx}
\usepackage{multirow}
\usepackage{tikz}
\usepackage[caption=false]{subfig}
\usepackage{relsize}	
\usepackage{array}
\usepackage{float}
\usepackage{color}
\usepackage{xcolor}
\usepackage{soul}
\usepackage{cellspace} 
\usepackage{ulem}

\usepackage{hyperref}
\hypersetup{colorlinks=true,linkcolor=blue,anchorcolor=blue,citecolor=red, filecolor=blue,urlcolor=blue,bookmarksnumbered=true,
pdfview=FitB
}

\def\dd{{\mathrm{d}}}

\mathchardef\-="2D

\colorlet{darkgreen}{green!60!black}
\colorlet{brightyellow}{yellow!75!red}
\colorlet{orange}{red!50!yellow}
\colorlet{darkblue}{blue!60!black}
\colorlet{darkred}{red!80!black}
\colorlet{greenblue}{green!50!blue}

\makeatletter

\newcommand{\Rmnum}[1]{\expandafter\@slowromancap\romannumeral #1@}
\makeatother

\begin{document}
\title{Relativistic dynamics of charmonia in strong magnetic fields}

\author{Liuyuan Wen}
\affiliation{Department of Modern Physics, University of Science \& Technology of China, Hefei 230026, China}

\author{Meijian Li}
\affiliation{Instituto Galego de Fisica de Altas Enerxias (IGFAE), Universidade de Santiago de Compostela, E-15782 Galicia, Spain}

\author{Yiyu Zhou}
\affiliation{Department of Physics, University of Turin, via Pietro Giuria 1, I-10125 Torino, Italy}
\affiliation{INFN, Section of Turin, via Pietro Giuria 1, I-10125 Torino, Italy}

\author{Yang Li}
\affiliation{Department of Modern Physics, University of Science \& Technology of China, Hefei 230026, China}
\affiliation{Anhui Center for Fundamental Sciences in Theoretical Physics, University of Science and Technology of China, Hefei 230026, China}

\author{James P. Vary}
\affiliation{Department of Physics and Astronomy, Iowa State University, Ames, IA 50011, USA}

\date{\today}
 \begin{abstract}
We investigate the properties of charmonium systems in strong external magnetic fields using a relativistic light-front Hamiltonian approach within the Basis Light-Front Quantization (BLFQ) framework. By solving the eigenvalue problem for the invariant mass squared operator with confinement potentials and one-gluon-exchange interactions, we obtain the mass spectrum and wave functions under varying magnetic fields. Our results reveal significant spectral modifications via the Zeeman effect, including $\eta_c$-$J/\psi$ mixing and magnetic sublevel splitting. Momentum density analysis demonstrates wave function deformation, with transverse momentum broadening and longitudinal narrowing under strong fields, alongside structural shifts in parton distributions such as double-hump profiles in excited states. Relativistic corrections and center-of-mass coupling critically drive these dynamics, highlighting the necessity of a relativistic framework for QCD bound states in extreme magnetic environments.
 \end{abstract}

\maketitle

\section{Introduction}
\label{sec:intro}

The properties of bound states in an external magnetic field represent one of the oldest problems in quantum mechanics. Throughout the previous century, this topic was extensively explored in both atomic and solid-state physics, leading to a rich variety of physics~\cite{Born:1989}. The dynamics of atoms can be significantly simplified due to several factors. First, the non-relativistic approximation is highly accurate for nearly all relevant systems. Second, the nuclei are much heavier than the electrons, meaning that the motion of the nucleus contributes only as a minor correction to the atomic spectrum. Third, magnetic fields achievable in the laboratories with $e B \lesssim 10^{-7} \mathrm{eV}^{2}$, are far smaller than the typical binding energies in atoms. Consequently, the effects of the external magnetic field on atoms and molecules remain a tiny perturbation. 

However, these approximations are not particularly satisfied when describing hadrons in the strong magnetic fields generated during relativistic heavy-ion collisions. For example, at RHIC in Brookhaven, the magnetic field produced during such collisions is estimated to reach values on the order of $\sim m_{\pi}^{2}$, where $m_{\pi}=140 \, \mathrm{MeV}$ is the pion mass. At LHC in CERN, the magnetic field can be even stronger, reaching $\sim 10 m_{\pi}^{2}$~\cite{Tuchin:2013ie, Bzdak:2011yy, Skokov:2009qp, Voronyuk:2011jd, Deng:2012pc, Huang:2015oca}. These intense magnetic fields may significantly alter the structure of QCD bound states. Furthermore, the masses of constituent quarks are neither sufficiently different to justify the neglecting of the center-of-mass (c.m.) motion, nor large enough to validate the non-relativistic approximation in the super-intense magnetic fields created during heavy-ion collisions. Despite these limitations, these assumptions continue to serve as the foundation for most, if not all, research in the field, owing to their great phenomenological importance~\cite{Marasinghe:2011bt, Machado:2013rta, Alford:2013jva,Cho:2014exa,Cho:2014loa, Simonov:2013jpa,Simonov:2012if,Rougemont:2014efa,Dudal:2014jfa,Guo:2015nsa, Andreichikov:2013zba, Bonati:2015dka,Hattori:2015aki, Gubler:2015qok,Suzuki:2016kcs, Yoshida:2016xgm, Andreichikov:2016ayj, Singh:2017nfa,Hasan:2017fmf, Liu:2018zag, Iwasaki:2018pby, Hattori:2019ijy, Yakhshiev:2019gvb, Mishra:2020kts,Hasan:2020iwa,Chen:2020xsr,S:2021qkj,Kirscher:2021zuj,Khan:2021syq, Chuang:2021vgl,Nilima:2022tmz,De:2022gse,Lal:2023ebr,Sebastian:2023tlw,Mishra:2023uhx,Chen:2024lmp,Hattori:2023egw,Endrodi:2024cqn}. In particular, non-relativistic potential models have been widely utilized~\cite{Godfrey:1985xj}, as described by the Hamiltonian,
\begin{equation}\label{eqn:NRQM_Hamiltonian}
H=\sum_{i} \frac{\left(\vec{p}_{i}-q_{i} \vec{A}\right)^{2}}{2 m_{i}}+q_{i} \phi-\vec{\mu}_{i} \cdot \vec{B}+m_{i}+\sum_{i, j} V_{i j}
\end{equation}
Within this framework, various systems have been investigated, including S-wave quarkonia~\cite{Alford:2013jva}, excited S-wave quarkonia, heavy-light mesons~\cite{Suzuki:2016kcs,Yoshida:2016xgm}, and P-wave quarkonia~\cite{Bonati:2015dka,Iwasaki:2018pby}. Other approaches employed to address the problem include holographic QCD \cite{Iatrakis:2015sua, Jena:2022nzw, Li:2016gfn}, effective Lagrangians \cite{Amal:2018qln}, QCD sum rules \cite{Machado:2013rta,Cho:2014exa,Cho:2014loa,Kumar:2019tiw, Parui:2022msu}, and lattice QCD~\cite{Endrodi:2024cqn}.

Non-relativistic QCD (NRQCD) estimates put the velocity of the charm quark within charmonium as $v^2/c^2 \sim 0.3$, which is mildly relativistic. Indeed, NRQCD calculation of the radiative transitions of charmonium encounters poor convergence even up to next-to-next-to-leading order (NNLO)\cite{Feng:2015uha, Feng:2017hlu}. 
A fully consistent treatment of the problem first necessitates a fully relativistic description of hadrons in vacuum. Recently, strides have been made in the light-front Hamiltonian approach to the hadron spectrum and hadron structures using a non-perturbative framework known as basis light-front quantization (BLFQ)~\cite{Vary:2009gt}. In this method, an effective light-front Hamiltonian is constructed based on insights from AdS/QCD and the Bloch-Wilson renormalization of the QCD Hamiltonian quantized at a fixed light-front time $x^{+}=t+z/c$. The long-distance confinement physics of the system is captured by the former, while the short-distance one-gluon-exchange interaction is accounted for by the latter. The effective Hamiltonian is then diagonalized to yield the mass spectrum and wave functions which are further used to compute hadronic observables. This approach has been successfully applied to describe the quarkonium spectrum in vacuum~\cite{ Li:2015zda, Li:2017mlw, Tang:2019gvn, Tang:2020org, Qian:2020utg}, with the obtained spectra agreeing with experimental values within 40 MeV. Moreover, the resulting wave functions have been utilized to compute physical observables, such as leptonic and radiative widths, as well as the corresponding off-shell form factors~\cite{Li:2018uif, Adhikari:2018umb,Lan:2019img,  Li:2021ejv, Wang:2023nhb, Hu:2024edc, Xu:2024hfx}. These results show reasonable agreement with experimental measurements whenever such data are available.

In addition to providing a rigorous treatment of relativistic dynamics, a major advantage of light-front dynamics lies in its ability to directly access partonic observables, such as parton distribution functions. Furthermore, as a Hamiltonian-based method, the light-front Hamiltonian approach can be naturally extended to address time-dependent problems~\cite{Zhao:2013jia, Zhao:2013cma}. Examples include scattering processes in strong laser fields~\cite{Hu:2019hjx, Lei:2022nsk} and interactions with classical non-Abelian gauge~\cite{Li:2020uhl, Li:2021zaw, Li:2023jeh, Li:2025wzq}, the latter of which have also been implemented through quantum simulations \cite{Barata:2022wim, Barata:2023clv, Wu:2024adk, Qian:2024gph,Li:2024ufq}. This versatility makes the light-front Hamiltonian approach a powerful tool for exploring a wide range of phenomena in high-energy physics.

In this work, we extend this formalism to study the $c\bar c$ system in the presence of a strong uniform external magnetic field, which is highly relevant for heavy-ion physics. The introduction of the magnetic field breaks the rotational symmetry and lifts the degeneracy between states with different magnetic quantum numbers $m_j$, a phenomenon known as the Zeeman effect. As the strength of the magnetic field $B$ increases, states with different spins, such as $\eta_c$ and $J/\psi$, begin to mix. Additionally, the c.m. motion starts to couple with the intrinsic motion of the quarks and antiquarks, necessitating an updated framework for state identification. At even higher magnetic field strengths, the bound states are eventually ionized due to the Lorentz force. Furthermore, the inter-quark potential is modified by the external field, leading to a lowering of the open-flavor threshold. Across all these regimes, the structure of hadrons undergoes significant modifications, giving rise to physical signals that may be experimentally observable. For recent reviews on this topic, see Refs.~ \cite{Iwasaki:2021nrz, Rothkopf:2019ipj, Huang:2015oca, Mocsy:2013syh, Hattori:2023egw}.

The remainder of this article is organized as follows. In Sec.~\ref{sec:lfqcd}, we introduce the light-front Hamiltonian formalism, viz. light-front QCD coupled to an external magnetic field. In Sec.~\ref{sec:blfq}, we present the basis function method used in our calculations. The numerical results are presented and discussed in Sec.~\ref{sec:results}. Finally, we summarize our findings and conclude in Sec.~\ref{sec:conclusions}.

\section{Light-front QCD within an external magnetic field}
\label{sec:lfqcd}

\subsection{Coupling to the external magnetic fields}\label{sec:minimal_coupling}

The Lagrangian of the system in the presence of a strong classical external electromagnetic field $\mathcal{F}^{\alpha \beta}$ can be derived through minimal coupling, where the covariant derivative is modified as $D^{\mu} \rightarrow \mathcal{D}^{\mu} \equiv D^{\mu} + i e_f \mathcal{A}^\mu$. The resulting Lagrangian takes the form:
\begin{equation}\label{eqn:Lagrangian}
	\mathscr{L} = \mathscr{L}_{\textsc{ym}} + \bar{\psi}(i \slashed{\mathcal{D}} - m) \psi = \mathscr{L}_{\mathrm{QCD}} - J^\mu \mathcal{A}_\mu,
\end{equation}
where $D = \partial + i g_s G$ is the covariant derivative involving the gluon field $G$, and $\mathcal{A}$ represents the vector potential of the classical external field, viz. $\mathcal{F}^{\alpha \beta} = \partial^\alpha \mathcal{A}^\beta - \partial^\beta \mathcal{A}^\alpha$. Here, $e_f$ denotes the quark charge, and $J^\mu = \bar{\psi} \gamma^\mu \psi$ is the electromagnetic 4-current. 
The Lagrangian is then quantized on the light front at $x^+ = 0$ using the standard quantization procedure~\cite{Brodsky:1997de}. In this framework, the light-front time is defined as $x^+ = x^0 + x^3$, while the spatial coordinates in light-front variables are expressed as $x^- = x^0 - x^3$ and $\vec{x}_\perp = (x^1, x^2)$. For further details regarding the conventions employed in this work, see Appendix A of Ref.~\cite{Li:2017mlw}.

We adopt the light-cone gauge for the external field $\mathcal{A}^{+}=0$ as well as for the gluon field $G^{+}=0$. The resulting light-front QCD Hamiltonian within a classical external electromagnetic field is given by~\cite{Neville:1971uc, Ilderton:2013dba, Zhao:2013cma}:
	\begin{multline}
		P^{-}=P_{\mathrm{QCD}}^{-}  
		+ e_{f} \int \mathrm{~d}^{3} x J^{\mu} \mathcal{A}_{\mu} 
		+ \frac{e_{f}^{2}}{2} \int \mathrm{~d}^{3} x \bar{\psi}\gamma^{\mu} \mathcal{A}_{\mu} \frac{\gamma^{+}}{i \partial^{+}} \gamma^{\nu} \mathcal{A}_{\nu} \psi, \\ 
		 +\frac{e_{f} g_{s}}{2} \int \mathrm{~d}^{3} x \bar{\psi}_{f} \gamma^{\mu} \mathcal{A}_{\mu} \frac{\gamma^{+}}{i \partial^{+}} \gamma^{\nu} G_{\nu} \psi+\frac{e_{f} g_{s}}{2} \int \mathrm{~d}^{3} x \bar{\psi} \gamma^{\mu} G_{\mu} \frac{\gamma^{+}}{i \partial^{+}} \gamma^{\nu} \mathcal{A}_{\nu} \psi. \label{eqn:Hamiltonian}
	\end{multline}
Here, the summation over color (not shown) and flavor $(f)$ is implied. The terms in the second line represent modifications to the effective quark potential induced by the external field. These terms are of higher order $O(\alpha_s \alpha_{\mathrm{em}})$, and will be neglected in the initial investigation. The last term in (\ref{eqn:Hamiltonian}) is a seagull term, whose significance will become clear later. Notably, this seagull term is present even in the absence of the gluon field.

In this work, we consider a uniform background magnetic field oriented in the $z$-direction, $\vec{B} = B \hat{z}$. As previously mentioned, we adopt the light-cone gauge $\mathcal{A}^+ = 0$. The remaining degrees of freedom (d.o.f.) of this static field can be fixed by employing the symmetric gauge $\vec{\mathcal{A}} = \frac{1}{2} \vec{B} \times \vec{r}$. Consequently, the full expression for the background gauge field is given by:
\begin{equation}
	\mathcal{A}^\pm = 0, \quad \vec{\mathcal{A}}_\perp = \frac{1}{2} B \hat{z} \times \vec{x}_\perp. \label{eqn:background_gauge_field}
\end{equation}

To quantize the quark field, we expand the field operator in terms of creation and annihilation operators at the initial time $x^+ = 0$:
\begin{equation}\label{eqn:free_field_expansion}
	\psi(x)=\sum_{s} \int \frac{\mathrm{~d}^{3} p}{(2 \pi)^{3} 2 p^{+}}\Big[u_{s}(p) e^{-i p \cdot x} b_{s}(p) 
	+ v_{s}(p) e^{i p \cdot x} d_{s}^{\dagger}(p)\Big].
\end{equation}
Here, $b$ and $d$ represent the annihilation operators for the quark and antiquark, respectively. Using this free-field expansion, the Hamiltonian can be expressed in a second-quantized form. The second-quantized light-front Hamiltonian is more intuitively understood in its quantum many-body representation (see Appendix~\ref{eqn:secquant} for a detailed derivation):
\begin{equation}\label{eqn:LF_Hamiltonian_many-body}
	P^{-}=\sum_{i} \frac{(\vec{p}_{i \perp}-e_i\vec{\mathcal A}_{i\perp})^2+M_{i}^{2}}{p_{i}^{+}}- \vec{\mu}_{i} \cdot \vec{B} +V_{\mathrm{QCD}} .
\end{equation}
In this expression, $V_{\mathrm{QCD}}$ represents the QCD interaction, $M_i$ is the mass of th $i$-th parton, and $\vec{\mu}_i = g_S \left(e_i / p_i^+\right) \vec{S}_i$ represents the light-front magnetic moment, where the Landé $g$-factor $g_S = 2$ for free fermions, and $\vec{S}_i$ denotes the spinor matrix. This magnetic moment differs from the non-relativistic expression, as it does not involve the bare quark mass. 
Within the square term, it is straightforward to identify that the linear term $\vec{p}_{i\perp} \cdot \vec{\mathcal{A}}_{i\perp}$ originates from the minimal coupling term $J_\mu \mathcal{A}^\mu$ in Eq.~(\ref{eqn:Hamiltonian}), while the quadratic term $\vec{\mathcal{A}}_{i\perp}^2$ arises from the seagull term in Eq.~(\ref{eqn:Hamiltonian}). In principle, $V_{\mathrm{QCD}}$ also receives corrections due to the external magnetic field, as indicated in the Hamiltonian Eq.~(\ref{eqn:Hamiltonian}). However, for the purposes of the present applications, we neglect such contributions because they are small compared to the intrinsic QCD interactions.

In the symmetric gauge (\ref{eqn:background_gauge_field}), we further obtain:
\begin{equation}\label{eqn:LF_Hamiltonian_B}
	P^{-} = \sum_{i} \frac{\vec{p}_{i \perp}^{2} + M_{i}^{2}}{p_{i}^{+}} - \sum_{i} \left(\vec{m}_{i} + \vec{\mu}_{i}\right) \cdot \vec{B} + \frac{B^{2}}{4} \sum_{i} \frac{e_{i}^{2} \vec{r}_{i \perp}^{2}}{p_{i}^{+}} + V_{\mathrm{QCD}},
\end{equation}
where $\vec{m}_{i} = \left(g_{L} e_{i} / p_{i}^{+}\right) \vec{L}_{i}$ represents the magnetic moment associated with the orbital angular momentum $\vec{L}_{i} = \vec{r}_{i} \times \vec{p}_{i}$. The corresponding $g$-factor is $g_{L} = 1$.
Note that in light-front dynamics, the pair production terms $b^{\dagger} d^{\dagger}, d b$ vanish due to the Dirac delta function of longitudinal momentum conservation $\delta\left(p^{+}+p^{\prime+}\right)$. This is consistent with the observation that pair production in light-front dynamics happens only through zero modes~\cite{Hebenstreit:2011cr}.  

\subsection{Landau levels of a single quark}

Before investigating charmonium, let us first consider the single quark state $|q\rangle$ within an external magnetic field. The light-front Hamiltonian (\ref{eqn:LF_Hamiltonian_B}) becomes, 
\begin{equation}\label{eqn:single_quark_LF_Hamiltonian}
P^{-} = \frac{\vec{p}_{\perp}^{2} + M^{2}_q}{p^{+}} - \left(\vec{m} + \vec{\mu}\right) \cdot \vec{B} + \frac{B^{2}}{4} \frac{e_{q}^{2} \vec{r}_{\perp}^{2}}{p^{+}}. 
\end{equation}
The Schrödinger equation of this Hamiltonian is analytically solvable. The corresponding light-front energy is, 
\begin{equation}\label{eqn:Landau_levels}
p^- = \frac{M_q^2 + |e_qB|(2n+|m|+m+1)+2e_qB s_z}{p^+}, 
\end{equation}
where, $n=0,1, 2, \cdots; m=0,\pm 1, \pm2, \cdots; s_z = \pm \frac{1}{2}$. This is exactly the relativistic Landau levels. 

\subsection{Hadrons as one-particle states in vacuum}

According to Wigner classification, hadrons are the one-particle irreducible representation of the Poincaré group, which can be specified by a set of compatible operators from the Poincaré algebra given by \cite{Brodsky:1997de}:
\begin{equation}
	\left\{P^{+}, P^{-}, \vec{P}_{\perp}, \vec{\mathcal{J}}^{2}, \mathcal{J}_{z}, \mathrm{P}, \mathrm{C}\right\},
\end{equation}
where $\mathcal{J}^{2} = W^{2} / M^{2}$ represents the total intrinsic angular momentum of the system (i.e., the spin), and $W^{\mu}$ is the Pauli-Lubanski operator. The operator $\mathcal{J}_{z} = W^{+} / P^{+} = J_{z} - L_{z}$ corresponds to the projection of the total intrinsic angular momentum in the light-cone longitudinal direction. It differs from the total angular momentum $J_{z}$ by subtracting the orbital angular momentum $L_{z} = \vec{R}_{\perp} \times \vec{P}_{\perp}$ of the particle. Notably, while $\big[J_{z}, \vec{P}_{\perp}\big] = i \hat{z} \times \vec{P}_{\perp}$, we have $\big[\mathcal{J}_{z}, \vec{P}_{\perp}\big] = 0$. Here, $\mathrm{P}$ and $\mathrm{C}$ are the parity and charge conjugation operators, respectively.

Consequently, the hadronic state vector can be identified by the eigenvalues of these operators. Note that, for hadrons, the two Casimir operators, $\mathcal{M}^{2} = P_\mu P^\mu$ and $\vec{\mathcal{J}}^{2}$, label the intrinsic quantum numbers, the hadronic mass $M$ and the hadronic spin $j$, respectively. Operator $\mathcal M^2$ is also referred to as the light-cone Hamiltonian, denoted as $H_\textsc{lc} = P^{+} P^{-} - \vec{P}_{\perp}^2$. The corresponding eigenvalue equations are:
	\begin{equation}
		\begin{split}
			& H_\textsc{lc} |P^+, \vec P_\perp, M, j, m_j, \mathrm{P}, \mathrm{C}\rangle = M^2 |P^+, \vec P_\perp, M, j, m_j, \mathrm{P}, \mathrm{C}\rangle, \\
			& {\vec{\mathcal{J}}^2} |P^+, \vec P_\perp, M, j, m_j, \mathrm{P}, \mathrm{C}\rangle = j(j+1) |P^+, \vec P_\perp, M, j, m_j, \mathrm{P}, \mathrm{C}\rangle. \\
		\end{split}
	\end{equation}
%
Other operators can be constructed kinematically, allowing for the selection of more convenient sets of compatible operators. In BLFQ, the following set of compatible operators is adopted \cite{Li:2013cga}:
\begin{equation}
	\left\{P^{+}, H_\textsc{lc}, H_{\textsc{cm}}, L_{z}, \vec{\mathcal{J}}^{2}, \mathcal{J}_{z}, m_P, \mathrm{C}\right\},
\end{equation}
where $m_P$ represents the mirror parity (also known as the reflection symmetry), which flips only one spatial coordinate, such as $x^{1}$. Its relation with parity $\mathrm{P}$ is $m_P = (-i)^{2j}\mathrm{P}$. The operator $H_{\textsc{cm}} = \vec{P}_{\perp}^{2} + \big(P^{+} \Omega\big)^{2} \vec{R}_{\perp}^{2}$ is a harmonic oscillator Hamiltonian for the c.m.~motion and $\Omega$ is its oscillator frequency, and $L_{z} = \big(\vec{R}_{\perp} \times \vec{P}_{\perp}\big) \cdot \hat{z}$ corresponds to the projection of the orbital angular momentum of the c.m.~along the longitudinal direction. Here, $\vec{R}_{\perp}$ is the transverse c.m. coordinate, defined as:
\begin{equation}
	\vec{R}_{\perp} = \sum_{i} x_{i} \vec{r}_{i \perp}.
\end{equation}
Within this formalism, the c.m.~motion of a one-particle state is classified by the eigenvalues of these operators:
	\begin{equation}
		\begin{split}
			&  H_\textsc{cm} |P^+, n, m, M, j, m_j, m_P, \mathrm{C}\rangle = 2P^+\Omega(2n+|m|+1) |P^+, n, m, M, j, m_j, m_P, \mathrm{C}\rangle, \\
			&  L_z |P^+, n, m, M, j, m_j, m_P, \mathrm{C}\rangle = m |P^+, n, m, M, j, m_j, m_P, \mathrm{C}\rangle. \\
		\end{split}
	\end{equation}
These operators act only on the c.m.~part of the wave function and do not affect its intrinsic structure. Consequently, the wave function factorizes into intrinsic and c.m. components. 
The mass spectrum and wave functions are obtained by diagonalizing the generalized light-cone Hamiltonian:
\begin{multline}
	H_{\textsc{lc}} |P^+, n, m, M, j, m_j, m_P, \mathrm{C}\rangle  = \Big[ M^2 + 2\lambda (2n+|m|+1)P^+\Omega \Big] |P^+, n, m, M, j, m_j, m_P, \mathrm{C}\rangle 
\end{multline}
where, 
\begin{equation}
	H_{\textsc{lc}} = P^{+} P^{-} - \vec{P}_{\perp}^{2} + \lambda H_{\textsc{cm}}. 
\end{equation}
Here  C-number Lagrange multiplier $\lambda$ controls the contributions of the c.m. motion.

\subsection{One-particle states in a magnetic field}

The presence of a uniform classical magnetic field explicitly breaks the Poincaré symmetry. Consequently, the total angular momentum operator $\vec{\mathcal{J}}^{2}$ no longer commutes with the Hamiltonian $P^{-}$, while $\mathcal{J}_{z}$ and $P^{+}$ remain conserved operators. The situation regarding translational symmetry in the transverse directions is more subtle. Although the system remains translationally invariant in the directions perpendicular to the uniform magnetic field, the generators of these translations are not the canonical momentum operator $\vec{P}$ or the kinetic momentum (aka. the mechanical momentum) $\vec{\Pi} = M \dot{\vec{R}} = \vec{P} - q \vec{A}$. Instead, the conserved generators of transverse translations are given by the pseudo-momentum $\vec{K}_{\perp} = \vec{P}_{\perp} + \sum_{i} e_{i} \vec{\mathcal{A}}_{i \perp}$. 
It is important to note that if the total charge $q \equiv \sum_{i} e_{i} \neq 0$, the components of the pseudo-momentum do not commute, i.e., $\left[K_{x}, K_{y}\right] \neq 0$. This non-commutativity is the origin of the Aharonov-Bohm effect. Furthermore, it can be shown that $\left[K_{i}, \mathcal{J}_{z}\right] \neq 0$, indicating that the pseudo-momentum and the longitudinal projection of the total angular momentum are incompatible observables. Namely, states with definite $J_z$ may not have definite $K_i$.

In an external magnetic field, a particle and its antiparticle gyrate in different orbits. As a result, the charge conjugation symmetry is broken. However, it can be shown that a combined parity $m_PC$ is still conserved. 
Given these considerations, a set of compatible operators in the presence of a magnetic field can be chosen as:
\begin{equation}
	\left\{P^{+}, H_\textsc{lc}, {J}_{z}, m_P\mathrm{C}\right\}.
\end{equation}
The Lagrange multiplier $\lambda$ has to be set to zero since the c.m.~motion does not factorize within the external magnetic fields. 
Here, we choose to identify the eigenstates with $J_z$ instead of the pseudo-momentum $K_i$, since it is more convenient to construct basis states with definite $m_j$ in BLFQ.
The one-particle state is obtained by diagonalizing the light-cone Hamiltonian:
\begin{equation}
	H_\textsc{lc} |P^+, M_\textsc{lc}, m_j, m_PC\rangle = M^2_\textsc{lc} |P^+, M_\textsc{lc}, m_j, m_PC\rangle
\end{equation}
Note that the eigenvalue $M_\textsc{lc}$ depends on the external magnetic field $B$. 
To find the meaning of $M_\textsc{lc}$, we consider charmonium as a one-particle state in an external magnetic field. The eigenvalue $M_\textsc{lc}$ of this particle is given by (\ref{eqn:single_quark_LF_Hamiltonian}):
\begin{equation}
\begin{split}
M^2_\textsc{lc} =\,& M^2 - g_{\mathcal{J}} q \vec{\mathcal{J}} \cdot \vec{B} + \frac{(q B)^2}{4} R_{\perp}^2, \\
=\,& M^2
\end{split}
\end{equation}
where $q = \sum_i e_i$ is the total charge of the meson, and $g_{\mathcal{J}}$ is the Landé $g$-factor. For quarkonium, the total charge vanishes ($q = 0$). 
Therefore, the quantity $M_\textsc{lc}$ can be interpreted as the mass of the mesons in the presence of the magnetic field. Note that this quantity depends on $B$. In vacuum, it reduces to the rest mass of the meson. 
This quantity provides crucial insights into how the meson spectrum is modified under the influence of the external field.

\section{BASIS LIGHT-FRONT QUANTIZATION}
\label{sec:blfq}

The light-front QCD interaction $V_{\mathrm{QCD}}$ is defined in the general Hilbert space, which is not particularly convenient for addressing bound-state problems. Ideally, we seek an effective Hamiltonian that operates within a model space. The leading component of this model space is the valence Fock sector $|q \bar{q}\rangle$. However, the non-perturbative nature of QCD at the hadronic scale makes the calculation of the effective interquark interaction particularly challenging.

Lattice simulations involving heavy quark sources have confirmed the traditional linear confining potential \cite{Bali:2000gf}. Since $M^2$ plays the role of the Hamiltonian in light-front dynamics, it has been argued that the confining potential takes a quadratic form~\cite{Trawinski:2014msa}. Further advancements in the holographic description of QCD have proposed a quadratic effective confining potential of the form $P^{+} P_{\textsc{conf} \perp}^{-} = \kappa^4 \zeta_{\perp}^2$, derived from the Regge trajectory \cite{Brodsky:2014yha}. Here, $\zeta_{\perp} = \sqrt{x(1-x)} r_{\perp}$ is holographically mapped to the fifth coordinate in anti-de Sitter space, and $\kappa$ represents the strength of the confining interaction.
For confinement in the longitudinal direction, we adopt the version introduced in~\cite{Li:2015zda, Li:2017mlw, Li:2021jqb, Li:2022izo}, which produces the same ground-state mass and wave function as the 't Hooft interaction while being analytically solvable: $P^{+} P_{\textsc{conf} \|}^{-} = \sigma^2 \partial_x \left(x(1-x) \partial_x\right)$, where $\sigma$ denotes the strength of the longitudinal confinement. In the non-relativistic limit, rotational symmetry requires that $\sigma = \kappa^2 / (M_q + M_{\bar{q}}) + O(M_{q, \bar{q}}^{-1})$.
Note that after adopting the phenomenological confining potentials relevant for low-energy resolution, the quark mass $M_q$ should be treated as the constituent quark mass. 
For the short-distance part of the interaction, we adopt the Bloch-Wilson renormalized one-gluon-exchange interaction~\cite{Wiecki:2014ola, Li:2015zda, Li:2017mlw, Tang:2018myz, Tang:2019gvn},
\begin{equation}
	V_{\textsc{oge}} \equiv P^{+} P_{\textsc{oge}}^{-} = -C_F \frac{\alpha_s(Q^2)}{Q^2} \bar{u}' \gamma_\mu u \bar{v} \gamma^\mu v',
\end{equation}
where $Q^2$ is the squared average 4-momentum transfer. 

The light-cone Hamiltonian for a meson in a strong magnetic field is then given by:
\begin{multline}
	H_\textsc{lc} =\sum_{i} \frac{\vec{k}_{i\perp}^{2}+M_{i}^{2}}{x_i}+\frac{B^{2}}{4} \sum_{i} \frac{e_{i}^{2} \vec{r}_{i \perp}^{2}}{x_{i}} 
	  -\sum_{i}\left(\frac{g_L e_i}{x_i} \vec L_i+\frac{g_S e_i}{x_i} \vec S_i\right) \cdot \vec{B} 
	  + \kappa^{4} \sum_{i} x_{i} \vec{r}_{i \perp}^{2} - \kappa^{4} \vec{R}_{\perp}^{2}  \\
	  - \sigma^{2} \sum_{i<j} \sum_{i<j} \partial_{x_{i}}\left(x_{i} x_{j} \partial_{x_{j}}\right) + V_{\textsc{oge}},
\end{multline}
where, $x_i = p^+_i/P^+$ is the longitudinal momentum fraction, and $\vec k_{i\perp} = \vec p_{i\perp} - x_i\vec P_\perp$ is the boost invariant relative transverse momentum. 
In the present work, we only consider the valence Fock sector $|c\bar c\rangle$ for charmonium. The light-cone Hamiltonian takes the form:
	\begin{align}
		& H_\textsc{lc} = \frac{\vec{k}_{\perp}^{2}+M_{c}^{2}}{x(1-x)}+\kappa^{4} x(1-x) \vec{r}_{\perp}^{2}-\frac{\kappa^{4}}{4 M_{c}^{2}} \partial_{x}\left(x(1-x) \partial_{x}\right)+V_{\textsc{oge}} \notag \\
		& \quad+\frac{\left(q_{f} e B\right)^{2}}{4}\left[\frac{r_{1 \perp}^{2}}{x}+\frac{r_{2 \perp}^{2}}{1-x}\right]-q_{f} e B\left(\frac{m_{1}+2 s_{1}}{x}-\frac{m_{2}+2 s_{2}}{1-x}\right).
	\end{align}
Here, $r_{1\perp}$ ($r_{2\perp}$) is the transverse coordinate of the quark (antiquark), and $\vec r_\perp = r_{1\perp} - r_{2\perp}$. $m_i = L_{iz} = (\vec r_{1\perp} \times \vec p_{1\perp})_z$ is the orbital angular momentum projected to the $z$ direction. $s_i = \pm 1/2$ is the spin projection. 

This eigenvalue equation can be solved using the basis function method. In contrast to the vacuum solution \cite{Li:2017mlw}, the interaction also affects the c.m.~motion. Therefore, it is necessary to extend the basis space to incorporate the c.m.~motion. 
The hadronic state vector can thus be represented as:
\begin{equation} \label{eqn:basis_space}
		|\psi\rangle  =\sum_{\substack{N, M \\ n, m, l, s_{1}, s_{2}}} \psi\big(N, M, n, m, l, s_{1}, s_{2}\big) \big|N, M, n, m, l, s_{1}, s_{2}\big\rangle ,
\end{equation}
where each basis state $|N, M, n, m, l, s_1, s_2\rangle \equiv |N, M\rangle_{\textsc{cm}} \otimes |n, m, l\rangle_{\textsc{rel}} \otimes |s_1, s_2\rangle_{\textsc{spin}}$ is a direct product of the c.m. motion, relative motion, and spin components. 
As in our previous work Ref.~\cite{Li:2017mlw}, we adopt the holographic wave functions as the basis, and the wave function $\psi_{s_1 s_2}(\vec{P}_\perp, x, \vec{k}_\perp) = \langle P^+, \vec{P}, k^+, \vec{k}_\perp | \psi \rangle$, where $x = k^+ / P^+$, can be represented as:
	\begin{equation}
		\psi_{s_{1} s_{2}}\big(\vec{P}_{\perp}, x, \vec{k}_{\perp}\big)  =\sum_{\substack{N, M \\
				n, m, l, s_{1}, s_{2}}} \psi\big(N, M, n, m, l, s_{1}, s_{2}\big) \phi_{N M}\big(\vec{P}_{\perp}\big) \phi_{n m}\big(\vec{k}_{\perp} / \sqrt{x(1-x)} \big) \chi_{l}(x),
	\end{equation}
where, the holographic wave functions are defined as:
	\begin{align}
		& \phi_{n m}\left(\vec{p}_{\perp} ; b\right)=b^{-1} \sqrt{\frac{4 \pi n!}{(n+|m|)!}}\left(\frac{p_{\perp}}{b}\right)^{|m|} e^{-\frac{p_{\perp}^{2}}{2 b^{2}}} L_{n}^{|m|}\left(\frac{p_{\perp}^{2}}{b^{2}}\right) e^{i m \arg \vec{p}_{\perp}} \\
		& \chi_{l}(x ; \alpha, \beta)=\sqrt{\frac{4 \pi \Gamma(l+\alpha+1) \Gamma(l+\beta+1)}{\Gamma(l+\alpha+\beta+1) \Gamma(l+1)}} x^{\frac{\beta}{2}}(1-x)^{\frac{\alpha}{2}} P_{l}^{(\alpha, \beta)}(2 x-1).
	\end{align}
Here, $b = \kappa$, $\alpha = \beta = 2M_c / \sigma$, and $\beta = 2M_{\bar c} / \sigma$ are model parameters. We adopt the same values of the model parameters in vacuum as shown in Table~\ref{tab:model_parameters}. 

The matrix elements of the light-cone Hamiltonian $H_\textsc{lc}$ can be obtained from the integral:
	\begin{multline}\label{eqn:LC_Hamiltonian_matrix_elements}
		 \langle N', M'; l', n', m', s'_1, s'_2 | H_\textsc{lc} | N, M; l, n, m, s_1, s_2\rangle = \\
		 \int \frac{\dd x}{2x(1-x)} \int \frac{\dd^2k_\perp}{(2\pi)^3}
		\int \frac{\dd^2P_\perp}{(2\pi)^2}
		\int \frac{\dd x'}{2x'(1-x')} \int \frac{\dd^2k'_\perp}{(2\pi)^3}
		\int \frac{\dd^2P'_\perp}{(2\pi)^2} \\ 
		 \times \phi^*_{N'M'}(\vec P_\perp)\chi_{l'}(x')\phi_{n'm'}^*(\vec k'_\perp/\sqrt{x'(1-x')})
		  \phi_{NM}(\vec P_\perp)\chi_{l}(x)\phi_{nm}(\vec k_\perp/\sqrt{x(1-x)}) \\
		 \times \langle \vec P'_\perp; x', \vec k'_\perp, s'_1, s'_2| H_\textsc{lc} | \vec P_\perp; x, \vec k_\perp, s_1, s_2 \rangle. 
	\end{multline}
%
The basis is chosen such that the soft part of the Hamiltonian is diagonal:
	\begin{equation}
		\frac{\vec{k}_{\perp}^{2}+M_{c}^{2}}{x(1-x)}+\kappa^{4} x(1-x) \vec{r}_{\perp}^{2}+V_{\|}=2 \kappa^{2}\left(2 n+|m|+l+\frac{3}{2}\right)+4 M_{c}^{2}+\frac{\kappa^{4}}{4 M_{c}^{2}} l(l+1).
	\end{equation}
The matrix elements of the one-gluon-exchange interaction, $V_{\textsc{oge}}$, are computed in Ref.~\cite{Li:2017mlw}. The $B$-dependent terms are given by:
\begin{multline}
	\langle N', M'; l', n', m', s'_1, s'_2 | \Big(\frac{r_{1 \perp}^{2}}{x}+\frac{r_{2 \perp}^{2}}{1-x} \Big) | N, M; l, n, m, s_1, s_2\rangle = \\
	b^{-2}\delta_{s_1s'_1}\delta_{s_2s'_2}\delta_{M+m,M'+m'} \int \frac{\dd x}{4\pi} \chi_{l'}(x)\chi_{l}(x) 
	\Bigg\{ \\
	\frac{\delta_{MM'}\delta_{mm'}\delta_{nn'}}{x(1-x)} \Big[\delta_{NN'}(2N+|M|+1) 
	+\delta_{N,N'+1}\sqrt{N(N+|M|)} 
	+ \delta_{N',N+1}\sqrt{N'(N'+|M'|)}\Big] \\
	+ \frac{1-3x+3x^2}{x^2(1-x)^2}
	\delta_{MM'}\delta_{mm'}\delta_{NN'}\Big[ 
	\delta_{nn'}(2n+|m|+1)+\delta_{n,n'+1}\sqrt{n(n+|m|)} + \delta_{n',n+1}\sqrt{n'(n'+|m'|)}\Big] \\
	+ \frac{1-2x}{\sqrt{x^3(1-x)^3}} (-1)^{N+n-N'-n'+\frac{1}{2}(|M|+|m|-|M'|-|m'|)} \\ 
	\times \big(f_{NMN'M'}f_{n'm'nm}\delta_{M+1,M'}\delta_{m'+1,m} 
	+
	f_{N'M'NM}f_{nmn'm'}\delta_{M'+1,M}\delta_{m+1,m'}\big) 
	\Bigg\},
\end{multline}
and
\begin{multline}
	\langle N', M'; l', n', m', s'_1, s'_2 | \Big(\frac{m_{1}+2 s_{1}}{x}-\frac{m_{2}+2 s_{2}}{1-x}\Big) | N, M; l, n, m, s_1, s_2\rangle = \\
	\delta_{m+M,m'+M'}
	\delta_{s_1s'_1}\delta_{s_2s'_2} \Bigg\{ 
	\delta_{MM'}\delta_{nn'}\delta_{mm'}\int \dd x \, \chi_l(x)\chi_{l'}(x) \Big( \frac{m+2s_1}{x} - \frac{m+2s_2}{1-x}\Big)   \\
	- \int \dd x \, \chi_{l'}(x) \chi_l(x) \frac{1}{\sqrt{x(1-x)}}  
	\Big(\delta_{m',m+1} f_{nmn'm'}g_{NMN'M'}
	+\delta_{m,m'+1} f_{n'm'nm}g_{N'M'NM}\Big) \Bigg\},
\end{multline}
where $m_i = (\vec r_{i\perp} \times \vec p_{i\perp})_z$, and the functions $f_{nmn'm'}$ and $g_{nmn'm'}$ are defined as:
\begin{equation}
	f_{nmn'm'} = 
	\begin{cases}
		\sqrt{n+|m|+1}\delta_{n,n'} - \sqrt{n}\delta_{n,n'+1}, & (m\ge 0) \\
		\sqrt{n+|m|}\delta_{n,n'} - \sqrt{n'}\delta_{n',n+1}, & (m<0)
	\end{cases}
\end{equation}
and
\begin{equation}
	g_{nmn'm'} = 
	\begin{cases}
		- \sqrt{n+|m|+1}\delta_{n,n'} - \sqrt{n}\delta_{n,n'+1}, & (m\ge 0) \\
		\sqrt{n+|m|}\delta_{n,n'} + \sqrt{n'}\delta_{n',n+1}, & (m<0).
	\end{cases}
\end{equation}
Finally, the longitudinal integrals can be evaluated using the Jacobi-Gaussian quadrature.

Computers can only diagonalize finite-dimensional matrices. For practical numerical calculations, it is necessary to truncate the basis space. In the longitudinal direction, we restrict the quantum numbers up to $L_{\max}$. In the transverse direction, we adopt the $\bar N_{\max}$ truncation scheme, which limits the quantum numbers to a full shell:
\begin{align}
	& 2N + |M| + 1 + 2n + |m| + 1 \leq \bar N_{\max}, \\
	& 0 \leq l \leq L_{\max}, \\
	& M + m + s_1 + s_2 = m_j.
\end{align}
Note that, here, $\bar N_{\max}$-truncation incorporate the c.m. excitation including its corresponding zero-point energy. 
The last condition arises from angular momentum conservation. Once again, orbital angular momentum ($M$) due to the c.m.~motion is incorporated in $m_j$. Thus, our finite-truncated basis space is labeled by three regulators: $\mathcal{B}(\bar N_{\max}, L_{\max}, m_j)$. The continuum limit is achieved by extrapolating $\bar N_{\max} \to \infty$ and $L_{\max} \to \infty$. 
In our previous work Ref.~\cite{Li:2017mlw}, we tune the model parameters for each finite basis space $\mathcal{B}(\bar N_{\max}, L_{\max}, m_j=0)$ to reproduce the charmonium spectrum. We find that $N_{\max} = L_{\max} = 8$ suffices for reproducing the mass spectrum as well as the leptonic widths and the radiative widths. We thus adopt $\bar N_{\max} = 9, L_{\max} = 8$ as well as the same model parameters for the QCD part (see Table~\ref{tab:model_parameters}). And there is no new free parameter in the present work.

\begin{table}
	\centering
	\caption{Model parameters. $\kappa$ is the strength of the transverse confining potential, and $\sigma = \kappa^2/2M_c$ is the strength of the longitudinal confinement. $M_c$ is the charm quark mass. Note that $M_c$ is not the current charm quark mass.} $\bar N_{\max}$ and $L_{\max}$ are basis regulators. See the main texts for details. 
	\label{tab:model_parameters}
	\begin{tabular}{ScScScScScScSc}
		\toprule 
		$\kappa$  & $M_c$ & $\sigma$ & $\bar N_{\max}$ & $L_{\max}$ \\ 
		\hline 
		0.98 GeV &  1.57 GeV & 0.31 GeV & 9 & 8 \\ 
		\botrule 
	\end{tabular}
\end{table}


\section{Numerical results}
\label{sec:results}

\subsection{Mass spectrum}

\begin{figure}
	\centering
	\includegraphics[width=0.7\textwidth]{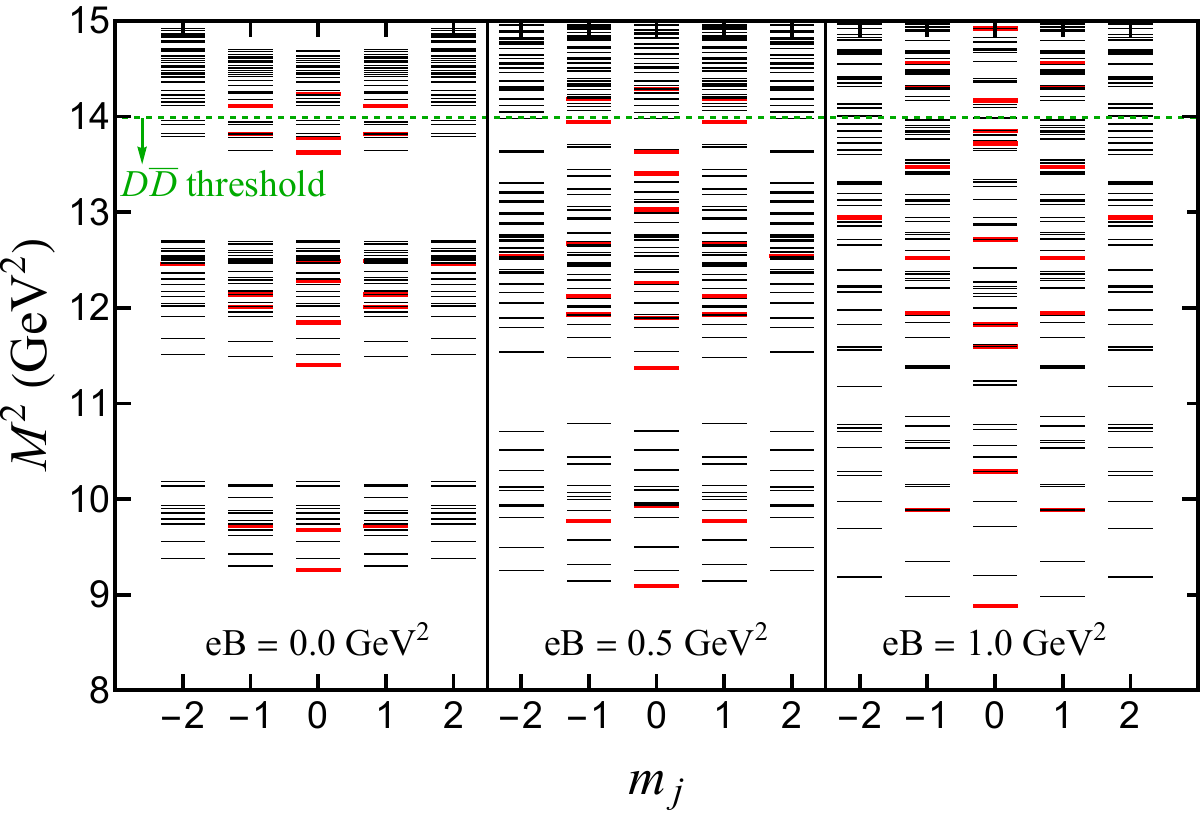}
	\caption{The levels obtained from diagonalizing the invariant mass squared $M^2$ with $eB = 0, 0.5, 1.0 \,\mathrm{GeV}^2$. Each set of levels are organized by the total magnetic projection $m_j$. For $eB=0$, we mark intrinsic levels, i.e. states without center of mass motion, with thick red bars, whereas the rest (thin black bars) are charmonium with c.m.~excitations. 
	For $eB\ne 0$, we use methods described below to identify the intrinsic states, and similarly mark them with thicker red bar.  }
	\label{fig:levels}
\end{figure}

The Hamiltonian matrix is constructed for different $m_j$ sectors according to Eq.~(\ref{eqn:LC_Hamiltonian_matrix_elements}). We diagonalize the Hamiltonian matrix using LAPACK \cite{lapack} to obtain the mass spectrum and wave functions. The data that support the findings of this article are openly available at Ref.~\cite{ref:data}. Figure~\ref{fig:levels} presents the mass squared levels for $eB = 0, 0.5, 1.0 \, \mathrm{GeV}^2$, corresponding to total magnetic projections $m_j = 0, \pm 1, \pm 2$. Note that the mass levels for $\pm m_j$ are degenerate, as required by the $m_P C$ symmetry. 

For the vacuum solution ($eB = 0$), the number of levels exceeds what was previously obtained in Ref.~\cite{Li:2017mlw}, as shown in the left panel of Fig.~\ref{fig:levels}. This discrepancy arises because we have expanded the basis space here to include the c.m.~motion. Consequently, states with c.m.~motion also appear in the low-lying spectrum. For instance, the ground states (g.s.) for $m_j = 0, \pm 1, \pm 2$ all correspond to $\eta_c$ but with different orbital angular momentum configurations. In Fig.~\ref{fig:levels}, their masses are approximately degenerate, as expected. Of course, the degeneracy is not exact due to the finite basis truncation. This highlights the need to identify states without c.m. motion, which we refer to as intrinsic states.
In BLFQ, the c.m. motion can be addressed by adding a Lagrange multiplier term to the Hamiltonian:
\begin{equation}
	H \quad \to \quad H + \lambda H_\textsc{cm},
\end{equation}
where $\lambda$ is a small positive number. The intrinsic states are then identified as those with minimal c.m.~excitation, characterized by $2N + |M| = 0$.
In Fig.~\ref{fig:levels}, the intrinsic mass levels for $eB = 0$ are highlighted with thick red bars in the left panel. All other states, indicated by thin black bars, correspond to excitations involving c.m. motion. We have verified that the spectrum of the intrinsic states is identical to our previous results obtained using only the relative basis.

For $eB \neq 0$, the Hamiltonian also acts on the c.m. motion. As a result, states with different c.m.~excitations are no longer degenerate, as is evident from the middle and right panels of Fig.~\ref{fig:levels}. As mentioned earlier, the combined parity $m_P C = (-1)^j PC$ remains an exact symmetry. Consequently, the spectrum can still be classified according to this quantum number. However, within each parity sector, different states, such as $\eta_c$ and $J/\psi$, mix with each other. Note that the orbital angular excitations also contribute to $m_P C$, leading to mixing between intrinsic states and states with c.m.~motion.

\begin{figure}
	\centering
	\includegraphics[width=0.48\textwidth]{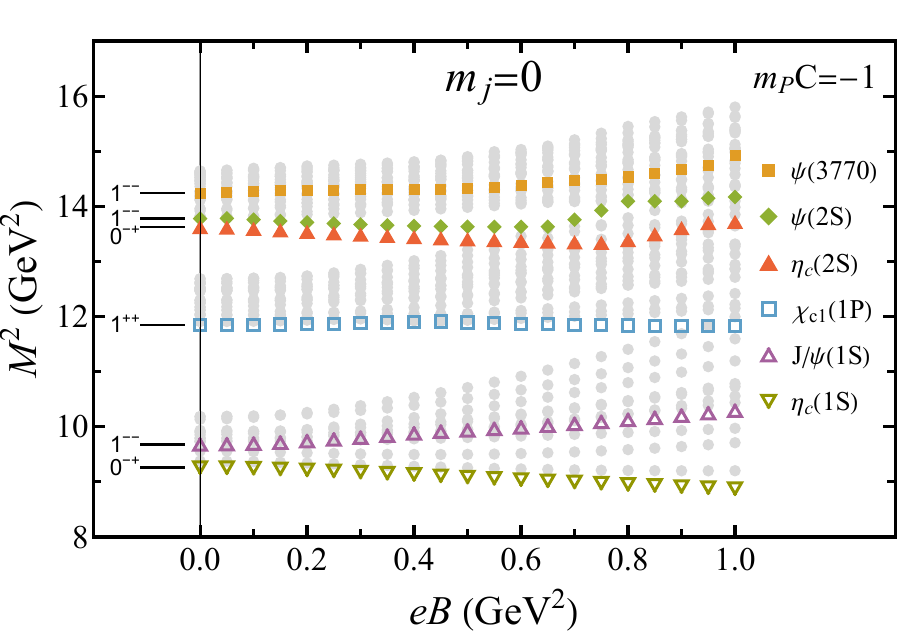}
	\includegraphics[width=0.48\textwidth]{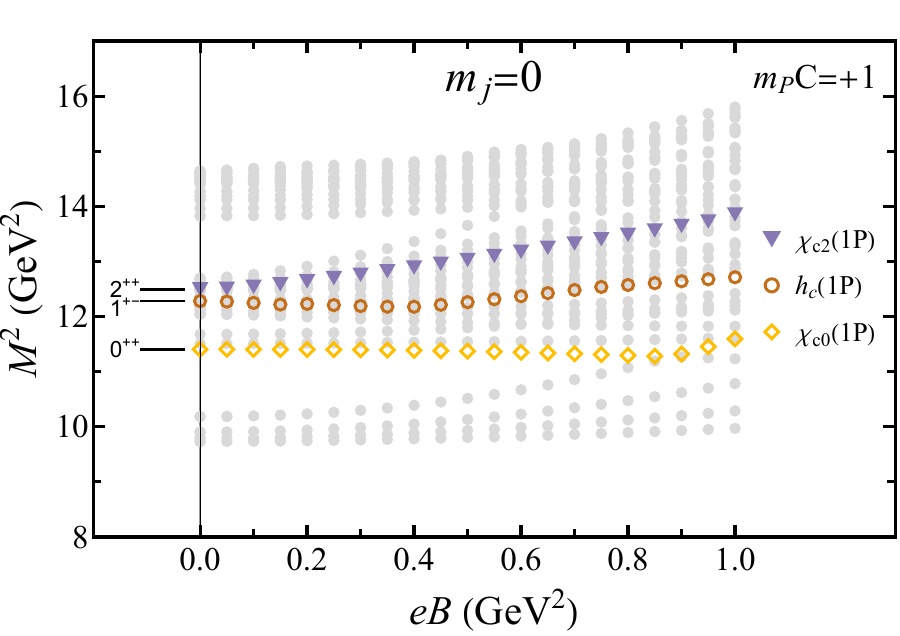}
	\includegraphics[width=0.48\textwidth]{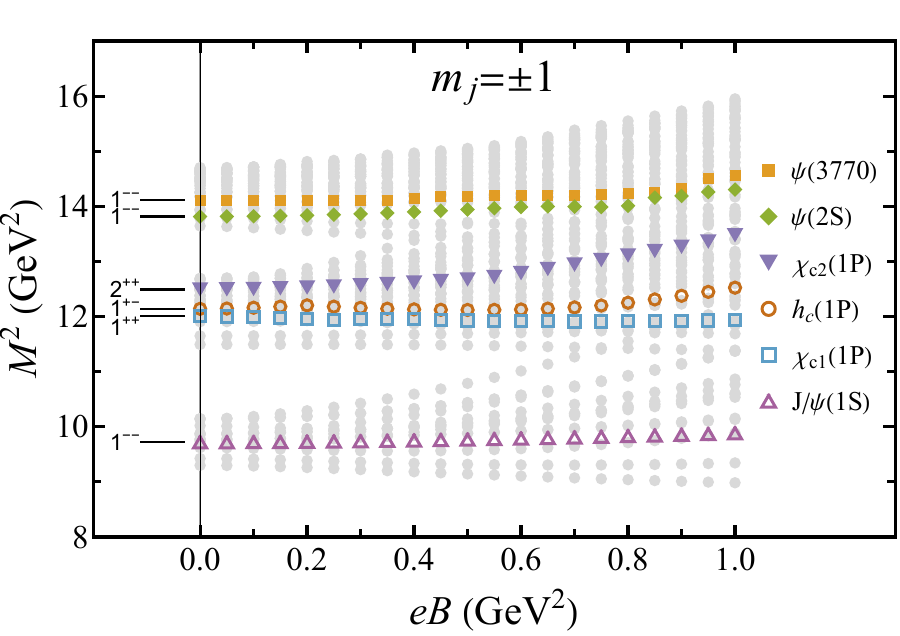}
	\includegraphics[width=0.48\textwidth]{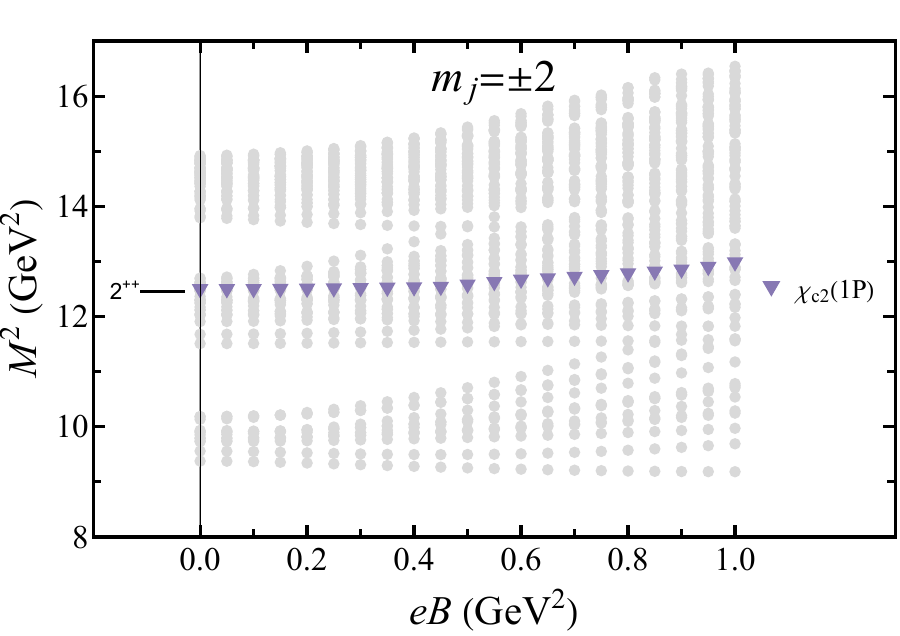}
	\caption{(Colors onlines) Charmonium levels as a function of $eB$ up to $eB = 1.0 \,\mathrm{GeV}^2$ for three different magnetic projections:  $m_j=0, \pm 1, \pm 2$. For $m_j=0$, the $m_PC = \pm 1$ sectors are shown separately. Charmonium levels in vacuum (black bars) are presented on the left for reference. We trace states that are connected to the intrinsic levels at $eB=0$ and label them using colorful symbols. These states are referred to as intrinsic states at $eB\ne 0$. The remaining levels, represented by gray dots, are states with center of mass excitations. 
	}
	\label{fig:scanB}
\end{figure}

On the other hand, for small and moderate values of $eB$, it is important to investigate how the intrinsic hadron structure is modified by the external magnetic fields. To identify intrinsic states in the presence of a magnetic field, we adopt the assumption that the mass of an intrinsic state varies smoothly as a function of $eB$. This assumption is valid in the absence of a phase transition, which is indeed the case for our two-body system in valence approximation. By contrast, states with c.m.~excitations exhibit mass gaps due to finite basis truncation, effectively acting as a large trap for the system.
Figure~\ref{fig:scanB} displays the charmonium states as functions of the magnetic field $eB$, ranging from $eB = 0$ to $eB = 1\,\mathrm{GeV}^2$, for three different magnetic projections: $m_j = 0, \pm 1, \pm 2$. The charmonium levels in vacuum (black bars) are shown on the left for reference. Specifically, we trace the trajectories connected to the intrinsic levels at $eB = 0$. These states are identified as intrinsic states and are marked with colorful symbols. To achieve this, we scan the spectrum with an increment of $e\Delta B = 2 \times 10^{-4}\,\mathrm{GeV}^2$, which is sufficient to resolve all fine structures within the trajectories. The remaining levels, represented by gray dots, correspond to states with c.m. excitations.
In practice, the identification of intrinsic states is further supported by examining the c.m. energy $2N + |M|$, which remains approximately zero, at least for small $eB$. Other quantum numbers -- exact or approximate-- such as the total spin $S$, mirror parity $m_P$, charge conjugation $C$, and the combined parity $m_P C$, are also useful for confirming the identification.
The above method for identifying intrinsic states is effective up to a moderate magnetic field strength of $eB \sim 1.0\,\mathrm{GeV}^2$. Beyond this range, the mixing of different charmonium states, as well as the mixing of c.m.~motion with intrinsic motion, complicates the identification process. For this reason, our identification terminates at $eB = 1.0\,\mathrm{GeV}^2$, as noted earlier.
The identified intrinsic states for $eB > 0$ are also highlighted with thick red bars in the middle and right panels of Fig.~\ref{fig:levels}.

 Our predicted $\eta_c$ and $J/\psi$ masses as functions of $eB$ exhibit trends similar to those reported in non-relativistic studies, such as Refs.~\cite{Alford:2013jva, Yoshida:2016xgm, Iwasaki:2018pby}. In particular, the hyperfine splitting between these two states increases with the strength of the magnetic field, a direct consequence of level repulsion. 
Table~\ref{table:level_mixing} illustrates the mixing between low-lying intrinsic charmonium states in vacuum ($eB = 0$) and in the presence of an external magnetic field with $eB = 0.1\,\mathrm{GeV}^2$ and $eB = 1.0\,\mathrm{GeV}^2$. As evident from the table, significant mixing between $\eta_c$ and $J/\psi$ occurs even at relatively small magnetic field strengths. As the magnetic field strength increases, the states along the trajectories of the intrinsic states become completely mixed with other states, including those involving c.m. excitations.

\begin{table}
\centering 
\caption{Mixing between low-lying intrinsic charmonia in vacuum $eB=0$ and in external magnetic field with $eB = 0.1, 1.0\,\mathrm{GeV}^2$.
Note that states with different combined parity $m_PC$ do not mix.}
\label{table:level_mixing}
\begin{tabular}{c|cc|ccc|ccc}
\toprule
& &  & \multicolumn{6}{c}{$eB=0.1\,\mathrm{GeV}^2$} \\ \cline{4-9}
& &  & \multicolumn{3}{c|}{$m_P C = -1$} & \multicolumn{3}{c}{$m_P C = +1$} \\ 
& $m_PC$ &  & $\eta_c$ & $J/\psi$ &  $\chi_{c1}$ & $\chi_{c0}$ & $h_c$ & $\chi_{c2}$ \\ \hline
\parbox[t]{2mm}{\multirow{6}{*}{\rotatebox[origin=c]{90}{$eB=0$}}} &
 \multirow{3}{*}{$-1$}  & $\eta_c$ &   ${0.98}$ & $0.18$ &  $-5.2\times 10^{-5}$ & 0 & 0 & 0 \\
&  & $J/\psi$ & $-0.18$ &  ${0.98}$  &  $-7.9\times 10^{-3}$ & 0 & 0 & 0 \\
& & $\chi_{c1}$ &  $-1.4\times 10^{-3}$ & $7.8\times 10^{-3}$ &   ${1.0}$ & 0 & 0 &  0\\ \cline{2-9}
& \multirow{3}{*}{$+1$} & $\chi_{c0}$ & 0 & 0 &   0 & {1.0} & $-0.071$ & $0.022$ \\
&  & $h_c$ &  0 & 0 &  0 & $0.075$ & ${0.92}$ &  $-0.38$ \\
& & $\chi_{c2}$ & 0 & 0 &   0 & $6.6\times 10^{-3}$ & $0.38$& $0.92$ \\ \hline
& &  & \multicolumn{6}{c}{$eB=1.0\,\mathrm{GeV}^2$} \\ \cline{4-9}
& &  & \multicolumn{3}{c|}{$m_P C = -1$} & \multicolumn{3}{c}{$m_P C = +1$} \\ 
& $m_PC$ &  & $\eta_c$ & $J/\psi$ &  $\chi_{c1}$ & $\chi_{c0}$ & $h_c$ & $\chi_{c2}$ \\ \hline
\parbox[t]{2mm}{\multirow{6}{*}{\rotatebox[origin=c]{90}{$eB=0$}}} &
 \multirow{3}{*}{$-1$}  & $\eta_c$ &   ${0.72}$ & $2.6\times 10^{-5}$ &  $4.7\times 10^{-5}$ & 0 & 0 & 0 \\
&  & $J/\psi$ & $-0.56$ &  $1.2\times 10^{-4}$  &  $2.7\times 10^{-4}$ & 0 & 0 & 0 \\
& & $\chi_{c1}$ &  $-0.041$ & $-4.6\times 10^{-5}$ &   ${-4.1}\times 10^{-3}$ & 0 & 0 &  0\\ \cline{2-9}
& \multirow{3}{*}{$+1$} & $\chi_{c0}$ & 0 & 0 &   0 &  $2.1\times 10^{-3}$ & $-0.011$ & $-1.5\times 10^{-3}$\\
&  & $h_c$ &  0 & 0 &  0 & $1.5\times 10^{-3}$ & ${8.1\times 10^{-3}}$ &  $6.4\times 10^{-3}$ \\
& & $\chi_{c2}$ & 0 & 0 &   0 & $8.0\times 10^{-4}$ & $0.014$& $-8.4\times 10^{-3}$ \\
\botrule
\end{tabular}
\end{table}
 
\begin{figure}[t]
	\centering
	\includegraphics[width=0.7\textwidth]{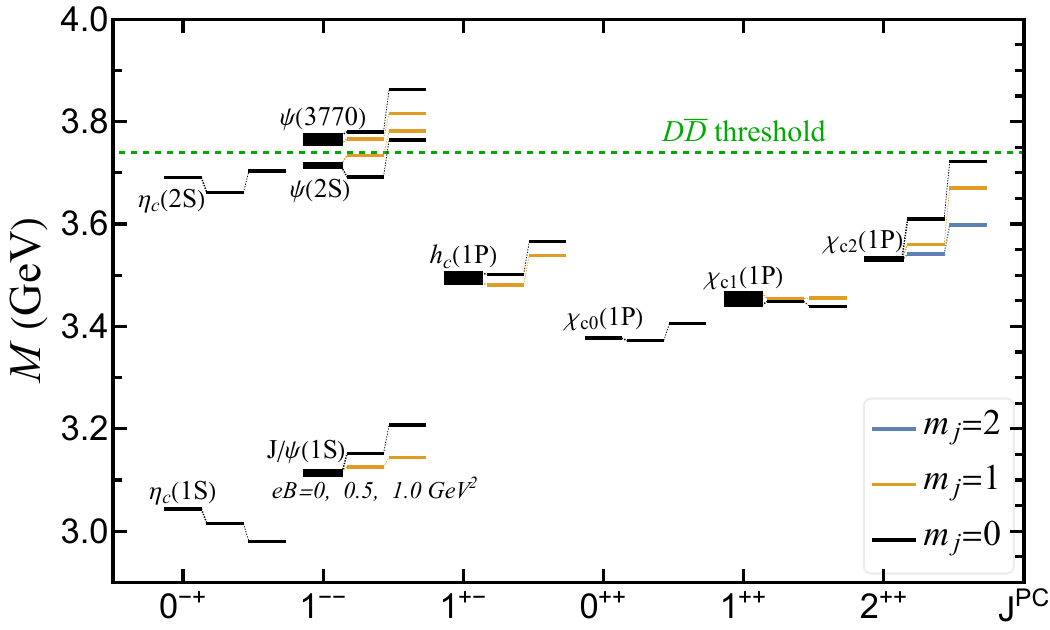}
	\caption{(Colors online) Spectrum of the intrinsic charmonium states within vacuum $eB = 0$ and within an external magnetic field at $eB = 0.5, 1.0\,\mathrm{GeV^2}$. For $eB \ne 0$, we use different color to represents states with different magnetic projection $m_j$. The splitting of the charmonium mass for different magnetic projections within the external magnetic field is known as the Zeeman effect. 
	In our calculation, the splitting of the magnetic projections still exists at $eB=0$, as caused by the violation of the exact rotation symmetry by the model truncation. We thus use black boxes to represent the spread of the mass eigenvalues for $eB=0$. 	}
	\label{fig:zeeman}
\end{figure}

Figure~\ref{fig:zeeman} displays the spectrum of intrinsic charmonium states in the presence of an external magnetic field with strengths $eB = 0, 0.5, 1.0\,\mathrm{GeV}^2$. For $eB = 0$, as previously mentioned, the degeneracy between states with different magnetic projections $m_j$ is only approximate. Following our earlier work \cite{Li:2017mlw}, we use boxes to represent the spread of the mass eigenvalues.
For $eB \neq 0$, states with different magnetic projections $m_j$ are represented by distinct colors. The splitting of charmonium masses for different $m_j$ values within the external magnetic field is known as the Zeeman effect. One should note that the Zeeman effect in quarkonium differs fundamentally from that in atoms. For instance, at small magnetic fields ($eB \ll \Lambda_\textsc{qcd}^2$), the level splitting $\Delta M^2$ is not linearly proportional to $B$. This is because the magnetic moment of quarkonium vanishes due to charge conjugation symmetry. This behavior is also reflected in the slopes of the charmonium masses $M^2$ as functions of $eB$. From Fig.~\ref{fig:scanB}, it is evident that these slopes vanish in the zero magnetic field limit. Thus, the Zeeman effect in quarkonium reflects genuine changes in the internal structure of the states for different magnetic projections.

\subsection{Relative momentum distribution}

To gain more insight into the system, we investigate the distribution of the relative momentum, which is defined as, 
\begin{equation}\label{eqn:density}
	\rho(x, \vec k_\perp) = \sum_{s_1,s_2} \rho_{s_1s_2}(x, \vec k_\perp) = \sum_{s_1,s_2}\int \frac{\dd^2P_\perp}{(2\pi)^2} \big| \psi_{s_1s_2}(\vec P_\perp, x, \vec k_\perp) \big|^2,
\end{equation}
where, $(P^+, \vec P_\perp)$ are the total momentum, i.e. the momentum of the hadron; $x = p^+/P^+$ is the longitudinal momentum fraction of the quark and $\vec k_\perp = \vec p_\perp - x\vec P_\perp$ is the relative transverse momentum that is a Fourier conjugate to $\vec r_\perp$ the separation of the quark and antiquark. $\rho_{s_1s_2}$ is the spin dependent density. In particular, we investigate four spin components: $\rho_{\uparrow\downarrow - \downarrow\uparrow}$, $\rho_{\uparrow\downarrow + \downarrow\uparrow}$, $\rho_{\uparrow\uparrow}$ and $\rho_{\downarrow\downarrow}$, where, 
\begin{equation}
\psi_{\uparrow\downarrow\pm\downarrow\uparrow}(\vec P_\perp, x, \vec k_\perp) = 
\frac{1}{\sqrt{2}}\Big[ \psi_{\uparrow\downarrow}(\vec P_\perp, x, \vec k_\perp) \pm  
\psi_{\downarrow\uparrow}(\vec P_\perp, x, \vec k_\perp) \Big].
\end{equation}
The relative momentum density is normalized to unity, 
\begin{equation}
\int \frac{\dd x}{2x(1-x)}\int \frac{\dd^2 k_\perp}{(2\pi)^3} \rho(x, \vec k_\perp) = 1.
\end{equation}

\begin{figure}
	\centering
	\subfloat[\ g.s. of $m_PC=-1$ ]{\includegraphics[width=0.24\textwidth]{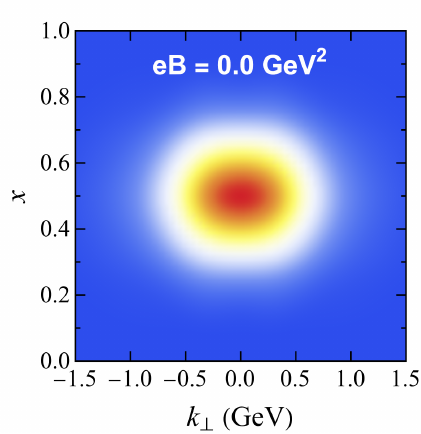}
	\includegraphics[width=0.24\textwidth]{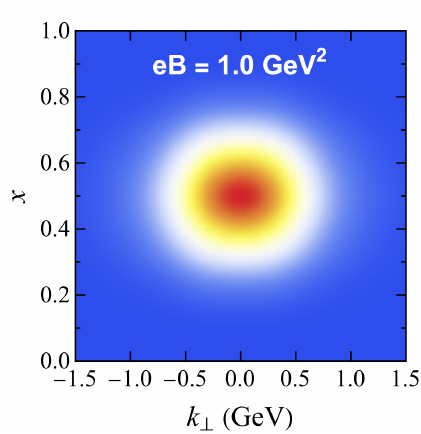}
	\includegraphics[width=0.24\textwidth]{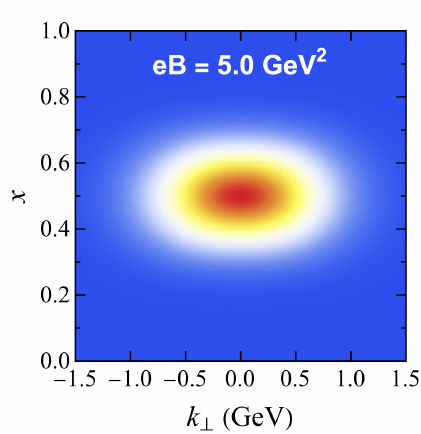}
	\includegraphics[width=0.24\textwidth]{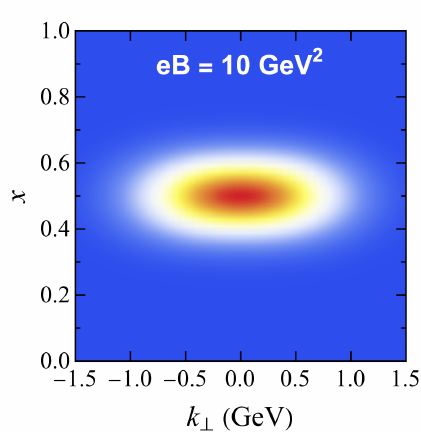}} 

	\subfloat[\ g.s. of $m_PC=+1$ ]{\includegraphics[width=0.24\textwidth]{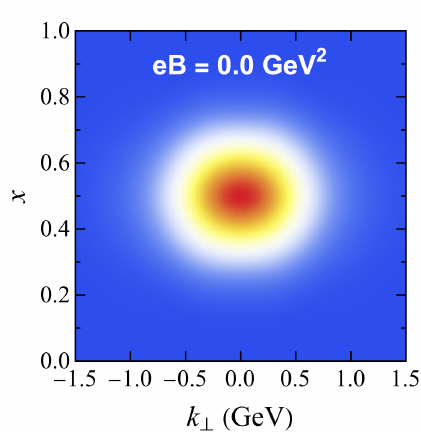}
	\includegraphics[width=0.24\textwidth]{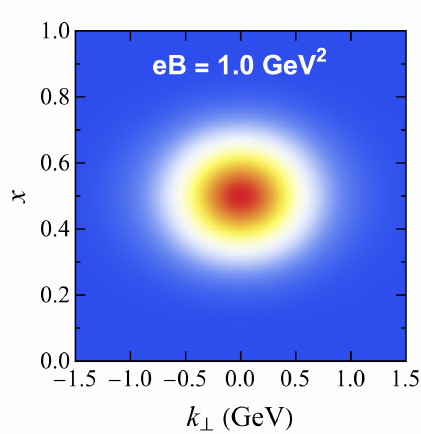}
	\includegraphics[width=0.24\textwidth]{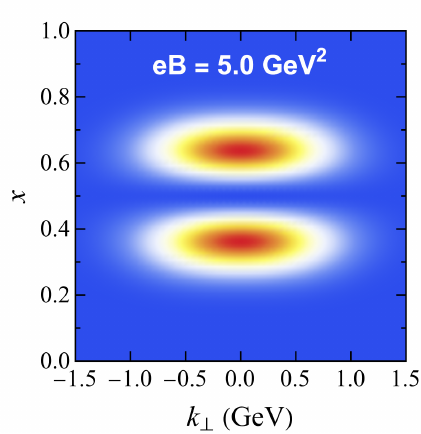}
	\includegraphics[width=0.24\textwidth]{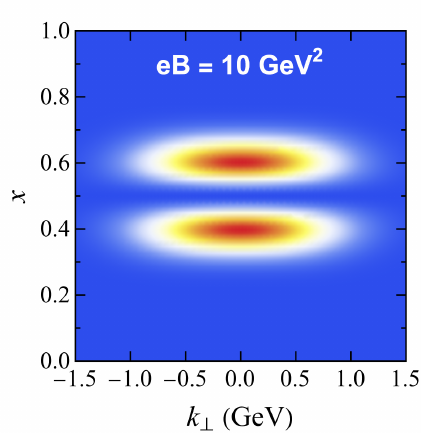}
	}
	\caption{
Comparison of the relative momentum density $\rho(x, k_\perp)$ of the ground state (g.s.) of within the $m_PC=\pm 1$ sectors in a magnetic field with $eB = 0, 1.0, 5.0, 10\,\mathrm{GeV}^2$. 
As the magnitude of magnetic field increases, both g.s. becomes more oblate, i.e. its longitudinal momentum distribution becomes narrower while its transverse momentum distribution becomes wider, which corresponds to a more prolate shape in coordinate space after a Fourier transform. 
For the  $m_PC=+1$ sector, the g.s.  at $eB \gtrsim 2.8\,\mathrm{GeV}^2$ becomes a longitudinal excited state. 
}	
\label{fig:gs_total_relative_momentum_density}
\end{figure}

\begin{figure}
	\centering
	\subfloat[$\uparrow \downarrow - \downarrow \uparrow$]{\includegraphics[width=0.24\textwidth]{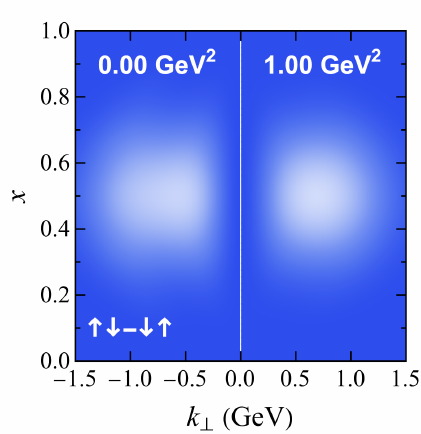}}\hspace{0.cm}
	\subfloat[$\uparrow \downarrow + \downarrow \uparrow$]{\includegraphics[width=0.24\textwidth]{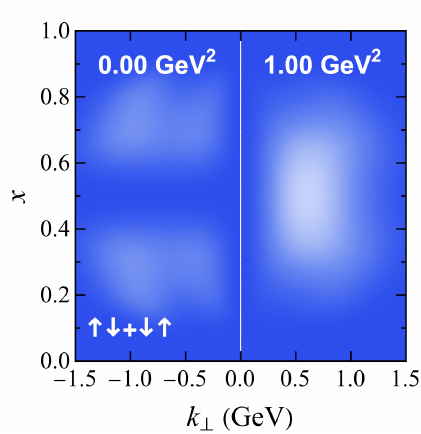}}\hspace{0.cm}
	\subfloat[$\uparrow \uparrow$]{\includegraphics[width=0.24\textwidth]{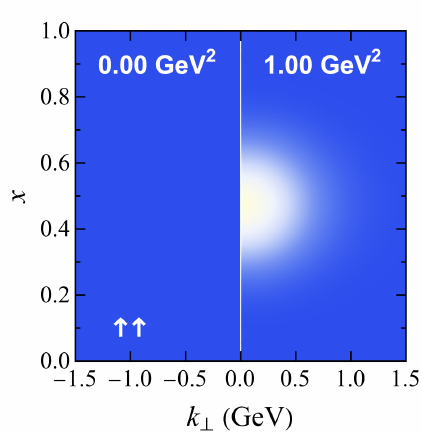}}\hspace{0.cm}
	\subfloat[$\downarrow \downarrow$]{\includegraphics[width=0.24\textwidth]{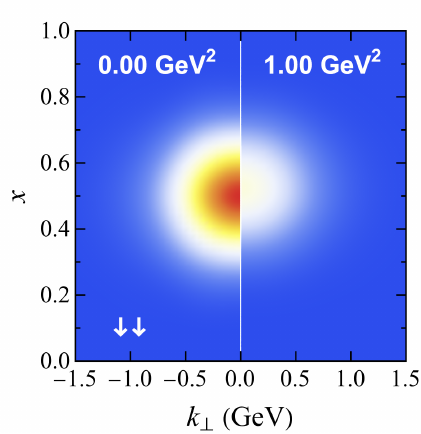}}
	\caption{
	Comparison of the spin dependent relative momentum densities of the g.s. charmonium with $m_PC=+1$ in vacuum $eB=0$ and in an external magnetic field $eB=1.0\,\mathrm{GeV}^2$. The spin structure of this state resembles $J/\psi$ with a c.m. orbital angular momentum $M=1$. As the external magnetic field turns on, the density $\psi_{\downarrow\downarrow}$ becomes asymmetric with respect to $x=0.5$, a consequence of the violation of the charge conjugation symmetry.}
	\label{fig:mpcp1_spin_momentum_density}
\end{figure}

Figure~\ref{fig:gs_total_relative_momentum_density} illustrates the relative momentum density $\rho(x, k_\perp)$ of the g.s. within the $m_PC = \pm 1$ sectors in a magnetic field with strengths $eB = 0, 1.0, 5.0, 10\,\mathrm{GeV}^2$.
At small $eB$, the g.s. of the $m_PC = -1$ sector is identified as the 1S pseudoscalar $\eta_c$ ($0^{-+}$). As we will demonstrate later in Fig.~\ref{fig:mpcp1_spin_momentum_density}, the g.s. of the $m_PC = +1$ sector corresponds to the 1S vector $J/\psi$ with an orbital angular excitation ($|M| = 1$). In both parity sectors, as the magnetic field strength increases, the g.s. becomes more oblate. Specifically, its longitudinal momentum distribution narrows, while its transverse momentum distribution broadens. This behavior translates to a more prolate shape in coordinate space after a Fourier transform, consistent with the classical expectation that a strong magnetic field restricts motion perpendicular to the magnetic field lines.
Furthermore, for the $m_PC = +1$ sector, the g.s. at $eB \gtrsim 2.8\,\mathrm{GeV}^2$ transitions into a longitudinally excited state ($l = 1, M = 0$), resembling the structure of $\chi_{c0}$ ($0^{++}$) due to the crossing of these two levels. Consequently, a double-hump structure appears above the transition, as we will also see later in Fig.~\ref{fig:pdf}.

\begin{figure}
	\centering
	\subfloat[$\uparrow \downarrow - \downarrow \uparrow$]{\includegraphics[width=0.24\textwidth]{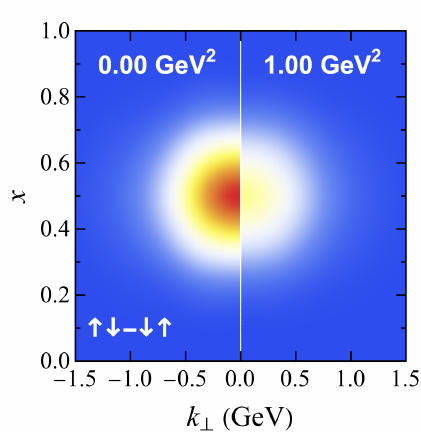}}\hspace{0.cm}
	\subfloat[$\uparrow \downarrow + \downarrow \uparrow$]{\includegraphics[width=0.24\textwidth]{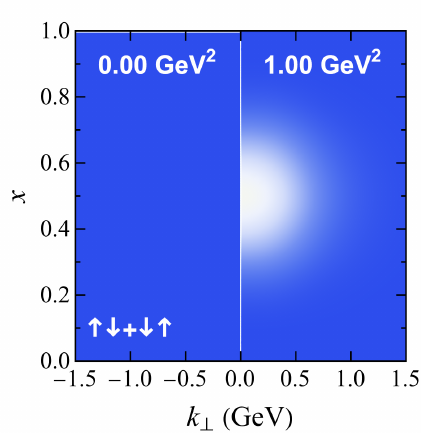}}\hspace{0.cm}
	\subfloat[$\uparrow \uparrow$]{\includegraphics[width=0.24\textwidth]{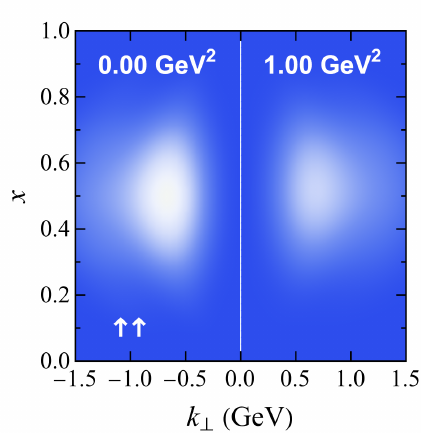}}\hspace{0.cm}
	\subfloat[$\downarrow \downarrow$]{\includegraphics[width=0.24\textwidth]{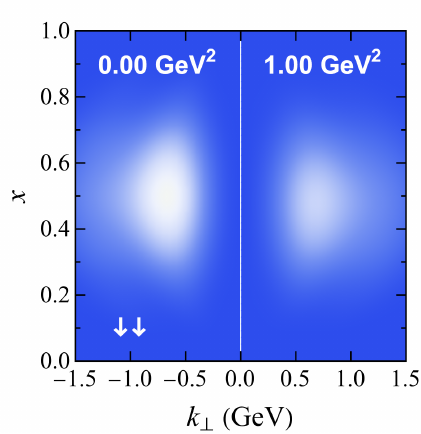}}
	\caption{
	Comparison of the spin dependent relative momentum densities of the g.s.~charmonium with $m_PC=-1$ in vacuum $eB=0$ and in an external magnetic field $eB=1.0\,\mathrm{GeV}^2$.  As the external magnetic field turns on, the parity forbidden spin triplet  component ($\psi_{\uparrow\downarrow+\downarrow\uparrow}$) becomes non-zero. Although it is not evident, the density becomes asymmetric with respect to $x=0.5$, a consequence of the violation of the charge conjugation symmetry.}
	\label{fig:mpcm1_spin_momentum_density}
\end{figure}

Figure~\ref{fig:mpcm1_spin_momentum_density} shows the spin dependent relative momentum densities of the g.s.~charmonium with $m_PC=-1$ in vacuum $eB=0$ and in an external magnetic field $eB=1.0\,\mathrm{GeV}^2$. As mentioned, this state is identified as the intrinsic state of $\eta_c$. As the external magnetic field turns on, the parity forbidden spin triplet  component ($\psi_{\uparrow\downarrow+\downarrow\uparrow}$) becomes non-zero, consistent with the state mixing observed in Table~\ref{table:level_mixing} within an external field. 
Although it is not evident, the density becomes asymmetric with respect to $x=0.5$, a consequence of the violation of the charge conjugation symmetry. 
Note that already in vacuum, $\eta_c$ developed spin-triplet wave functions $\psi_{\uparrow\uparrow} = \psi^*_{\downarrow\downarrow}$ due to relativistic effects \cite{Li:2017mlw, Li:2022mlg, Shi:2018zqd, Shi:2020pqe}.
Similarly, Fig.~\ref{fig:mpcp1_spin_momentum_density} shows the spin dependent relative momentum densities of the g.s.~charmonium with $m_PC=+1$ in vacuum $eB=0$ and in an external magnetic field $eB=1.0\,\mathrm{GeV}^2$. As mentioned, this state is identified as $J/\psi$ with orbital angular excitation. 
The momentum distribution also becomes asymmetric with respect to $x=0.5$, a consequence of the violation of the charge conjugation symmetry.

\begin{figure}
	\centering
	\subfloat[\ $m_PC=-1$]{\includegraphics[width=0.24\textwidth]{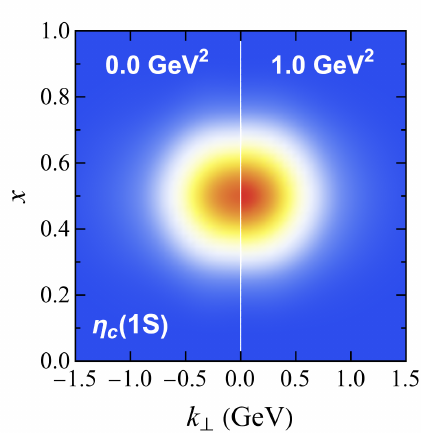}
	\includegraphics[width=0.24\textwidth]{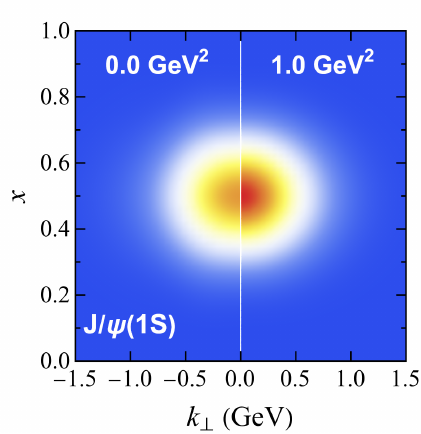}
	\includegraphics[width=0.24\textwidth]{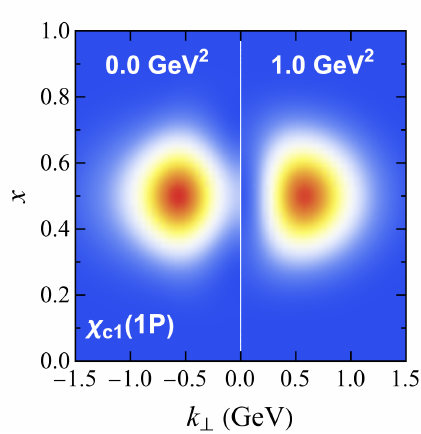}}
	
	\subfloat[\ $m_PC=+1$]{
	\includegraphics[width=0.24\textwidth]{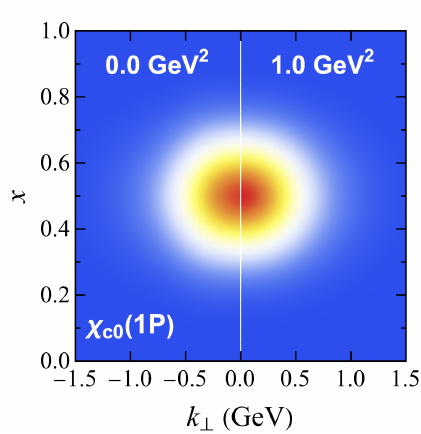}
	\includegraphics[width=0.24\textwidth]{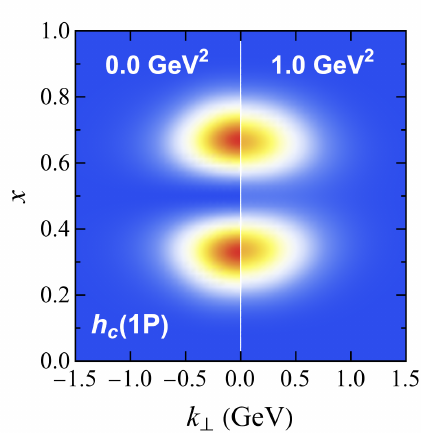}
	\includegraphics[width=0.24\textwidth]{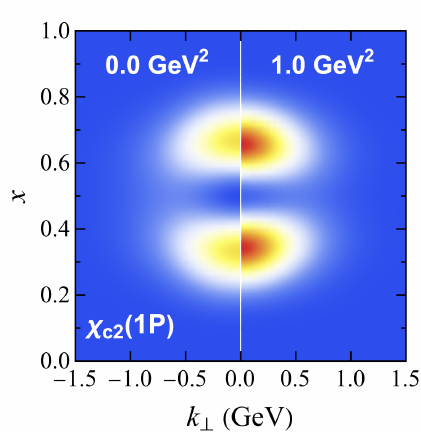}}
	\caption{
	Comparison of the relative momentum densities of several selected intrinsic states ($\eta_c, J/\psi, \chi_{c1}$ for $m_PC = -1$ and $\chi_{c0}, h_c, \chi_{c2}$ for $m_PC=+1$) at $eB = 0, 1.0\,\mathrm{GeV}^2$ with $m_j=0$. 
}	
	\label{fig:intrinsic_states_momentum_density}
\end{figure}

\begin{figure}
	\centering
	\subfloat[$m_j=-1$]{\includegraphics[width=0.24\textwidth]{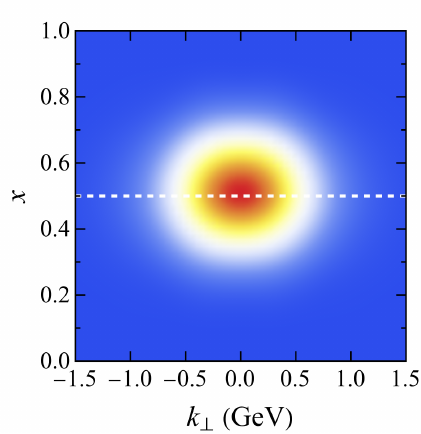}}\hspace{0.cm}
	\subfloat[$m_j=0$]{\includegraphics[width=0.24\textwidth]{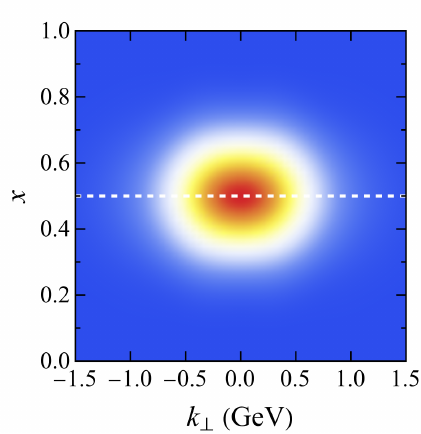}}\hspace{0.cm}
	\subfloat[$m_j=+1$]{\includegraphics[width=0.24\textwidth]{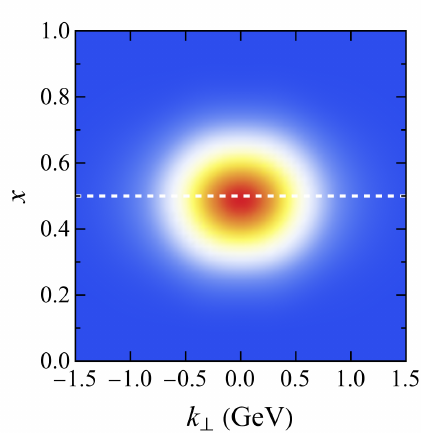}}\hspace{0.cm}
	\caption{Comparison of relative momentum density of $J/\psi$ for $m_j=0, \pm1$ at $eB = 0.4\,\mathrm{GeV}^2$. Note that the  relative momentum distributions of the polarized $J/\psi$ become asymmetric with respect to $x = 0.5$ due to the violation of the charge conjugation symmetry. The $m_j = +1$ and $m_j = -1$ states are related by the combined parity transformation $m_P C$.}
	\label{fig:Jpsi-mj-momentum}
\end{figure}

Figure~\ref{fig:intrinsic_states_momentum_density} compares the relative momentum densities of several selected intrinsic states ($\eta_c, J/\psi, \chi_{c1}$ for $m_PC = -1$ and $\chi_{c0}, h_c, \chi_{c2}$ for $m_PC=+1$) at $eB = 0$ and $eB = 1.0\,\mathrm{GeV}^2$, with $m_j=0$. Meanwhile, Fig.~\ref{fig:Jpsi-mj-momentum} displays the momentum densities of the $J/\psi$ in an external magnetic field $eB = 0.4\,\mathrm{GeV}^2$, for both longitudinal ($m_j=0$) and transverse ($m_j=\pm 1$) polarization. Beyond a slight deformation in shape, the density of $J/\psi$ shifts away from $x=0.5$ when the particle is transversely polarized in the external magnetic field. 
This shift originates from the interaction term $H_\textsc{int} = -\vec \mu_q \cdot \vec B -\vec \mu_{\bar q} \cdot \vec B$, where $\vec \mu_q = g_S (q_f e/x) \vec S_q$ and $\vec \mu_{\bar q} = - g_S (q_f e/(1-x)) \vec S_{\bar q}$ represent the light-front magnetic moments of the quark and antiquark, respectively. For $m_j=1$, the QCD interaction tends to align both the quark and antiquark spins parallel to $\vec B = B\hat z$. A decrease in $x$ reduces the interaction energy $H_\textsc{int}$, causing the wave function to shift toward smaller values of $x$.

These momentum distributions provide critical insights into the structure of the $c\bar c$ system under the influence of an external magnetic field. Through these densities, the effects of the magnetic field become more apparent, including the deformation of the system, the mixing of different quantum states, and the interplay between intrinsic motion and the c.m.~motion. Such observations highlight the intricate dynamics induced by the magnetic field and offer a deeper understanding of the underlying physics governing the charmonium system.

\subsection{Parton Distributions}
Parton distribution functions (PDFs) provide a detailed description of the longitudinal momentum fraction carried by the constituent partons -- in this case, the charm quark and antiquark -- within a hadronic bound state. In the light-front framework, the unpolarized PDF of a charmonium state can be obtained by integrating out the transverse momentum degrees of freedom and summing over all possible spin configurations. Mathematically, this is expressed as:
\begin{align}
	f(x) &= \frac{1}{2 x (1 - x)} \sum_{s_1, s_2} 
	\int \frac{\mathrm{d}^2 P_{\perp}}{(2 \pi)^2} \int \frac{\mathrm{d}^2 k_{\perp}}{(2 \pi)^2} 
	\bigl|\psi_{s_1 s_2}(\vec{P}_\perp, x, \vec{k}_\perp)\bigr|^2, \\
	&= \frac{1}{2 x (1 - x)} \int \frac{\mathrm{d}^2 k_{\perp}}{(2 \pi)^2}\,\rho\bigl(x,\vec{k}_\perp\bigr),
\end{align}
The quantity $\rho(x, \vec{k}_\perp)$ is the relative momentum density defined earlier in Eq.~(\ref{eqn:density}). 

\begin{figure}
	\centering
	\subfloat[\ g.s. of $m_PC=-1$]{\includegraphics[width=0.4\textwidth]{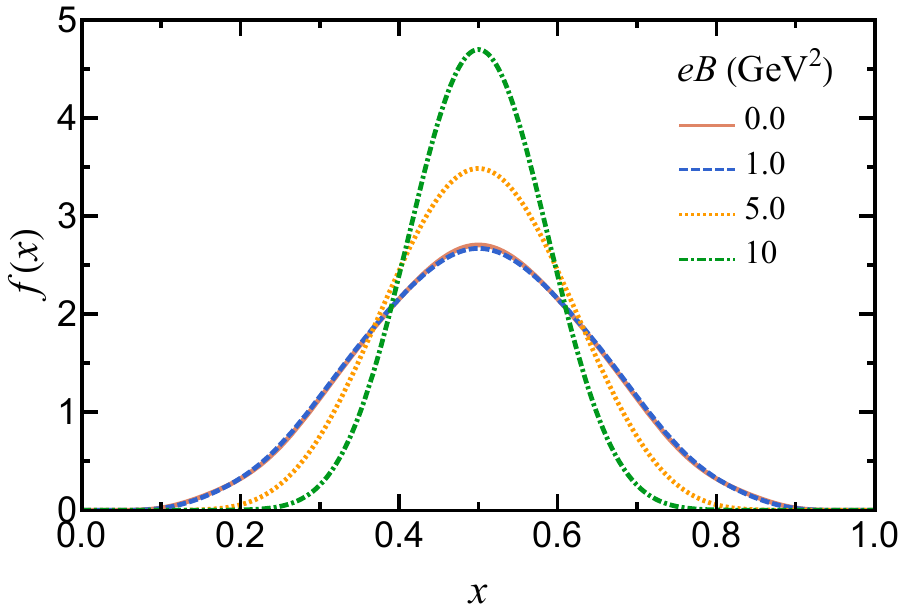}}\quad
	\subfloat[\ g.s. of $m_PC=+1$]{\includegraphics[width=0.4\textwidth]{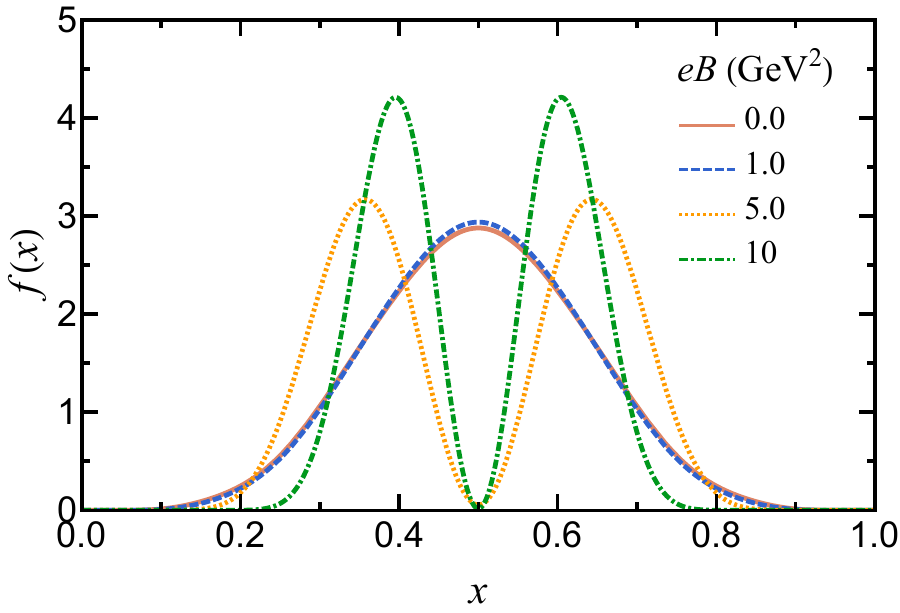}}
	\caption{Parton distribution functions of the g.s.~charmonia within external magnetic fields with $eB=0, 1.0, 5.0, 10 \, \mathrm{GeV}^2$ for parity sector $m_PC=-1$ (\text{Left}) and $m_PC=+1$ (\text{Right}). 
	This g.s.~of the $m_PC=-1$ sector is identified as $\eta_c$.  At small $eB$, the g.s.~of the $m_PC=+1$ sector is identified with $J/\psi$, also an S-wave. 
	As the strength of the magnetic field increases, the PDFs become narrower. For the  $m_PC=+1$ sector, the g.s.~PDF changes to a double-hump shape above the critical magnetic field $eB \approx 2.8 \,\mathrm{GeV}^2$. 
	}
	\label{fig:pdf}
\end{figure}

The resulting PDFs for the g.s.~in each parity sector ($m_PC = \pm 1$) are displayed in Fig.~\ref{fig:pdf}, with the magnetic field varying from $eB = 0$ to $eB = 10\,\mathrm{GeV}^2$. As previously discussed, the g.s. of the $m_PC = -1$ sector corresponds to the $\eta_c$, while at low magnetic field strengths ($eB$), the g.s. of the $m_PC = +1$ sector is identified as the $J/\psi$, which is also an S-wave state. These identifications account for the characteristic shapes of the g.s.~PDFs at small $eB$.
As the strength of the magnetic field increases, the PDFs become narrower. For the $m_PC=+1$ sector, the g.s.~PDF further changes to a double-hump shape above the critical magnetic field $eB \approx 2.8\,\mathrm{GeV}^2$, confirming the structural change shown in the momentum density distribution in Fig.~\ref{fig:gs_total_relative_momentum_density}.

\section{Discussion and conclusions}
\label{sec:conclusions}

In this work, we investigate the charmonium system in classical external magnetic fields within the basis light-front quantization (BLFQ) approach. 
By diagonalizing the light-cone Hamiltonian operator, we obtain the mass spectrum and light-front wave functions, and investigate the evolution of the charmonium structures as a function of $eB$ the strength of the external magnetic field. We found that the strong external magnetic fields significantly alter the spectral and structural properties of charmonium.  

At small to moderate field strength, we are able to identify intrinsic states and reconstruct the charmonium spectrum in $eB$. The spectrum shows a non-linear Zeeman splitting and level repulsion, which underscores the breakdown of rotation and parity symmetries. These effects are also shown in the analysis of the momentum densities. 
Under strong fields, transverse momentum broadening and longitudinal narrowing reflect the system's adaptation to magnetic confinement, aligning with classical expectations of restricted motion perpendicular to the field. However, the observed structural shifts -- such as double-hump profiles in parton distributions and the transition to longitudinally excited states in momentum densities -- reveal the competition between the strong force and the Lorentz force. These features emphasize the importance of center-of-mass coupling and orbital angular momentum excitations in shaping hadronic structure. 

One of the main features of the present work is the fully relativistic treatment of the dynamics in intense fields, which underlines the  significant mixing between intrinsic and c.m.~excitations, complicating the identification of physical states. This mixing, alongside the relativistic corrections to magnetic moments and confinement potentials, suggests that simplified models relying on static potentials or perturbative treatments may inadequately describe charmonium in heavy-ion collision environments. The framework developed here, validated by its consistency with experimental data in vacuum, provides a robust foundation for exploring transient phenomena in time-dependent fields, such as those encountered in quark-gluon plasmas.

While our valence-sector model captures essential QCD-electromagnetic interplay, the inclusion of explicit gluonic degrees of freedom (d.o.f.) allows us to investigate the modification of the strong force itself by the strong external magnetic field \cite{Lan:2021wok}.
Given the similarity between the Lorentz force and the Coriolis force, our work can be extended to the investigation of charmonium undergoing a rapid spinning \cite{Tuchin:2021yhy, Buzzegoli:2022omv}. Indeed, experimental measurements have revealed that quark-gluon plasma created in non-central relativistic heavy-ion collisions is the most vortical fluid known in nature \cite{Csernai:2013bqa, Csernai:2014ywa, Becattini:2015ska, Deng:2016gyh, Jiang:2016woz}. 
Another extension is the finite-temperature effects, which is needed to bridge the gap to experimental conditions. 
%

\acknowledgments

Y.L. and J.V. wish to extend their sincere gratitude to K. Tuchin for proposing the problem and engaging in discussions that laid the foundation for this work. 
This work is supported in part by the National Natural Science Foundation of China (NSFC) under Grant No.~12375081, and by the Chinese Academy of Sciences under Grant No.~YSBR-101.  
M.L. is supported by Xunta de Galicia under the ED431F 2023/10 project, European Union ERDF, the Spanish Research State Agency under project PID2020-119632GB-I00, and European Research Council under project ERC-2018-ADG-835105 YoctoLHC. 
Y.Z. is supported by the European Union ``Next Generation EU'' program through the Italian PRIN 2022 grant n. 20225ZHA7W.

\appendix

\section{Light-front Hamiltonian in second quantization}\label{eqn:secquant}

In this section, we provide a derivation of the quantum many-body Hamiltonian Eq.~(\ref{eqn:LF_Hamiltonian_many-body}) from the minimally coupled Lagrangian Eq.~(\ref{eqn:Lagrangian}). Our starting point is the light-front quantized Hamiltonian Eq.~(\ref{eqn:Hamiltonian}), 
\begin{equation} \label{eqn:Hamiltonian_v2}
		P^{-}= \int\mathrm{~d}^3x \, \overline\psi \gamma^+ \frac{-\nabla^2_\perp + m^2_f}{i\partial^+}\psi  + e_{f} \int \mathrm{~d}^{3} x \,  J^{\mu} \mathcal{A}_{\mu}+\frac{e_{f}^{2}}{2} \int \mathrm{~d}^{3} x \, \overline{\psi} \gamma^{\mu} \mathcal{A}_{\mu} \frac{\gamma^{+}}{i \partial^{+}} \gamma^{\nu} \mathcal{A}_{\nu} \psi + \cdots
\end{equation}
Here, the ellipsis represents terms involving gluon fields, including the QCD interactions as well as the interaction between gluons and the external fields $\mathcal A^\mu$ through instantaneous quarks. In our work, the external field only depends on $\vec x_\perp$. In fact, we adopt the symmetric gauge and the light-cone gauge: $\mathcal A^\pm = 0$, $\vec{\mathcal{A}}_\perp = (1/2)B \hat z \times \vec x_\perp$. $J^\mu = \overline\psi \gamma^\mu \psi$ is the current operator. Hereafter, we suppress the color indices, since the interaction with the external magnetic fields does not alter the color. Normal ordering is assumed. 

The kinetic term can be written in a second quantization form by substituting the free field expansion (\ref{eqn:free_field_expansion}), 
\begin{align}
P_0^- =\,& \int\mathrm{~d}^3x \, \overline\psi \gamma^+ \frac{-\nabla^2_\perp + m^2_f}{i\partial^+}\psi  \\
=\,& \sum_{s} \int\frac{\dd^3p}{(2\pi)^32p^+} \frac{p^2_\perp + m_f^2}{p^+}  \big[ b^\dagger_{s}(p)b_{s}(p) + d^\dagger_{s}(p)d_{s}(p) \big]\\
=\,& \sum_i \frac{p^2_{i\perp} + m_i^2}{p^+_i}
\end{align}
where, $b_{s}(p)$ [$d_{s}(p)$] is the quark (antiquark) annihilation operator. The quantum many-body expression in the last line follows from the correspondence between second quantization form and the quantum many-body form \cite{Sakurai:2011zz}.  The index $i$ in this line sums over particles in Fock space. 

The remaining two terms in (\ref{eqn:Hamiltonian_v2}) can also be written in a second quantization form by substituting the free field expansion (\ref{eqn:free_field_expansion}), 
\begin{align}
P^-_1 =\,& e_{f} \int \mathrm{~d}^{3} x \, J^{\mu} \mathcal{A}_{\mu} \\
=\,& e_f \int \dd^3x\, \sum_{s,s'}\int\frac{\dd^3 p}{(2\pi)^32p^+}\int\frac{\dd^3 p'}{(2\pi)^32p'^+} e^{i (p'-p)\cdot x} \\
& \times \Big\{
\frac{1}{2}\bar u_{s'}(p') \gamma^+ u_s(p) \mathcal A^-(x) - \bar u_{s'}(p') \vec \gamma_\perp u_s(p) \cdot \vec{\mathcal A}_{\perp}(x)
\Big\} 
b^\dagger_{s'}(p') b_s(p) + \text{anti-quark part}
\end{align}
Here, we focus on the quark part. The antiquark part is similar. The seagull term can be written as,
\begin{align}
P^-_2 =\,& \frac{e_{f}^{2}}{2} \int \mathrm{~d}^{3} x \,  \bar{\psi}\gamma^{\mu} \mathcal{A}_{\mu} \frac{\gamma^{+}}{i \partial^{+}} \gamma^{\nu} \mathcal{A}_{\nu} \psi \\
=\,& \frac{e_f^2}{2} \int \dd^3x\, \sum_{s,s'}\int\frac{\dd^3 p}{(2\pi)^32p^+}\int\frac{\dd^3 p'}{(2\pi)^32p'^+} e^{i (p'-p)\cdot x}  \bar u_{s'}(p') \big(\vec \gamma_\perp \cdot \vec{\mathcal A}_{\perp}\big)  \frac{\gamma^+}{p^+} \big(\vec\gamma_\perp\cdot \vec{\mathcal{A}}_\perp\big) u_s(p) \\
& \times b^\dagger_{s'}(p') b_s(p) + \text{anti-quark part}
\end{align}
The spinor matrix elements are, 
\begin{align}
& \bar u_{s'}(p') \gamma^+ u_s(p) = 2\sqrt{p^+p'^+}\delta_{ss'}, \\
& \bar u_{s'}(p') \vec \gamma_\perp u_s(p)\cdot\vec{\mathcal{A}}_\perp = (\vec p_\perp+\vec p'_\perp)\cdot\vec{\mathcal{A}}_\perp \delta_{ss'}+i 
\big[(\vec p_\perp - \vec p'_\perp)\times \vec{\mathcal{A}}_\perp\big] \cdot \vec \sigma_{ss'}, \\
&  \bar u_{s'}(p') \big(\vec \gamma_\perp \cdot \vec{\mathcal A}_{\perp}\big)  \gamma^+ \big(\vec\gamma_\perp\cdot \vec{\mathcal{A}}_\perp\big) u_s(p)
 = 2 \mathcal A_\perp^2 \delta_{ss'}
\end{align}
where, $\vec\sigma = (\sigma^1, \sigma^2, \sigma^3)$ are Pauli matrices. 
With these spinor identities, we obtain the second quantization form in momentum space, 
\begin{multline}
P^- = P^-_0 + P^-_1 + P^-_2 = 
\sum_{s} \int\frac{\dd^3p}{(2\pi)^32p^+} \frac{p^2_\perp + m_f^2}{p^+} b^\dagger_{s}(p)b_{s}(p) 
+ e_f\sum_{s, s'}\int \frac{\dd p^+}{(2\pi)2p^+}  \int \frac{\dd^2p_\perp}{(2\pi)^2} \int \frac{\dd^2p'_\perp}{(2\pi)^2} \\
\times \int \dd^2 x_\perp e^{i(\vec p'_\perp - \vec p_\perp)\cdot \vec x_\perp} \Big\{\mathcal A^-(x_\perp) \delta_{ss'} - \frac{(\vec p'_\perp +\vec p_\perp)\cdot\vec{\mathcal{A}}_\perp \delta_{ss'} + i\big[(\vec p_\perp - \vec p'_\perp)\times\vec{\mathcal{A}}_\perp\big]\cdot\vec\sigma_{ss'}}{p^+}  \Big\}b_{s'}^\dagger(p')b_s(p) \\
+ e_f^2 \sum_{s}\int \frac{\dd p^+}{(2\pi)2p^+}  \int \frac{\dd^2p_\perp}{(2\pi)^2} \int \frac{\dd^2p'_\perp}{(2\pi)^2} \int\dd^2x_\perp e^{i(\vec p'_\perp - \vec p_\perp)\cdot \vec x_\perp} \frac{\mathcal A_\perp^2}{p^+}b_{s'}^\dagger(p')b_s(p) + \text{antiquark part}
\end{multline}
In the above expression, we integrated out $x^-$ by taking advantage of the fact that $\mathcal A^\mu$ only depends on $\vec x_\perp$. 

Note that the transverse part is not diagonal. To diagonalize this part, we introduce the transverse coordinate-space representation, 
\begin{align}\label{eq:coordinate_space_operator}
B_s(p^+, \vec x_\perp) =\,&  \int\frac{\dd^2p_\perp}{(2\pi)^2} e^{-i\vec p_\perp \cdot \vec x_\perp} b_s(p), \\
b_s(p) =\,& \int \dd^2 x_\perp \, e^{i\vec p_\perp \cdot \vec x_\perp} B_s(p^+, \vec x_\perp)
\end{align}
where, the operators satisfy the canonical commutation relations, 
\begin{align}
\big\{ b_{s'}(p'), b^\dagger_s(p) \big\} =\,& 2p^+(2\pi)^3\delta^3(p-p')\delta_{ss'}, \\
\big\{ B_{s'}(p'^+, \vec x'_\perp), B^\dagger_s(p^+, \vec x_\perp)\big\} =\,& (2\pi)2p^+ \delta(p'^+ - p^+) \delta^2(x_\perp - x'_\perp).
\end{align}
By expressing the $b_s(p)$ in terms of the $B_s(p^+, \vec x_\perp)$ in Eq. (\ref{eq:coordinate_space_operator}), we obtain, 
\begin{multline}
P^- = \sum_s \int \frac{\dd p^+}{4\pi p^+} \int \dd^2 x_\perp 
\bigg\{
B^\dagger_s(p^+, \vec x_\perp) \Big[ e_f\mathcal A^- + \frac{-\nabla^2_\perp + m^2_f + e_f^2 \vec{\mathcal{A}}^2_\perp}{p^+} \Big] B_s(p^+, \vec x_\perp) \\
+ e_f \frac{B^\dagger_s(p^+, \vec x_\perp)\vec{\mathcal{A}}_\perp \cdot i\nabla_\perp B_s(p^+, \vec x_\perp) - \vec{\mathcal{A}}_\perp\cdot i\nabla_\perp B^\dagger_s(p^+, \vec x_\perp) B_s(p^+, \vec x_\perp)}{p^+} \bigg\} \\
+ \sum_{s,s'} \int \frac{\dd p^+}{4\pi p^+} \int \dd^2 x_\perp e_f \bigg\{ \frac{\Big[B^\dagger_{s'}(p^+, \vec x_\perp) \nabla_\perp B_s(p^+, \vec x_\perp) \times \vec{\mathcal{A}}_\perp \Big]\cdot \vec \sigma_{ss'}}{p^+} \\
+ \frac{\Big[ \nabla_\perp B^\dagger_{s'}(p^+, \vec x_\perp)  \times \vec{\mathcal{A}}_\perp B_{s}(p^+, \vec x_\perp)  \Big]\cdot \vec \sigma_{ss'}}{p^+} \bigg\}.
\end{multline}
The above expression can be written in a more compact form, 
\begin{multline}\label{eqn:second_quantized_Hamiltonian}
P^- = \sum_s \int \frac{\dd p^+}{4\pi p^+} \int \dd^2 x_\perp 
B^\dagger_s(p^+, \vec x_\perp) \Big[ e_f\mathcal A^- + \frac{(-i\nabla_\perp - e_f\vec{\mathcal{A}}_\perp)^2+ m^2_f}{p^+} \Big] B_s(p^+, \vec x_\perp) \\
- \sum_{s,s'} \int \frac{\dd p^+}{4\pi p^+} \int \dd^2 x_\perp B^\dagger_{s'}(p^+, \vec x_\perp) B_s(p^+, \vec x_\perp) \frac{e_f\big(\nabla_\perp\times \vec{\mathcal{A}}_\perp\big) \cdot \vec \sigma_{ss'}}{p^+}
\end{multline}
Note that the cross term of $B^\dagger_s(p^+, \vec x_\perp) (-i\nabla_\perp - e_f\vec{\mathcal{A}}_\perp)^2 B_s(p^+, \vec x_\perp)$ should be interpreted as 
\begin{equation}
B^\dagger_s(p^+, \vec x_\perp) i\nabla_\perp \cdot \big(\vec{\mathcal{A}}_\perp B_s(p^+, \vec x_\perp)\big) + B^\dagger_s(p^+, \vec x_\perp) \vec{\mathcal{A}}_\perp \cdot i\nabla_\perp B_s(p^+, \vec x_\perp),
\end{equation}
which is the same as that in the normal ordered expression in momentum space: $(\vec p_\perp - e_f \vec{\mathcal{A}}_\perp)^2 = \vec p^2_\perp -e_f \vec p_\perp \cdot \vec{\mathcal{A}}_\perp - e_f \vec{\mathcal{A}}_\perp\cdot \vec p_\perp + e_f^2\vec{\mathcal{A}}_\perp^2$. 

The second quantized Hamiltonian Eq.~(\ref{eqn:second_quantized_Hamiltonian}) is now diagonal and can be converted to the quantum many-body form:
\begin{equation}
P^- = \sum_i \frac{(\vec p_{i\perp} - e_i \vec{\mathcal A}_{i\perp})^2+m_i^2}{p^+_i} +  e_i \mathcal A^-_i - \frac{g_ie_i}{p^+_i} \vec B \cdot \vec S_i 
\end{equation}
where, $\mathcal A_i^\mu = \mathcal A^\mu(x_i)$, $\vec p_{i\perp} = -i\nabla_{i\perp}$,  $\vec B = \nabla \times \vec{\mathcal A}$, and $g_i = 2$,  $\vec S_i = \frac{1}{2} \vec\sigma_i$ for quarks. We thereby arrive at Eq.~(\ref{eqn:LF_Hamiltonian_many-body}).

\end{document}